\pdfoutput=1
\PassOptionsToPackage{hyphens}{url}\makeatletter\def\@classoptionslist{english}\makeatother%
\documentclass[conference]{IEEEtran}

\newcommand{\hidetodos}{}
\newcommand{\hidehighlights}{}

\usepackage[utf8]{inputenc}

\usepackage{ifluatex}
\ifluatex
\else
  \usepackage[T1]{fontenc}
\fi

\usepackage{xparse} %
\usepackage{ifthen}
\usepackage{calc}
\usepackage{relsize}
\usepackage{tabularx}
\usepackage{pgfplotstable}
\usepackage{subcaption}
\usepackage[inline]{enumitem}

\usepackage[main=english,ngerman]{babel}
\makeatletter
\def\l@en{\l@english}
\def\l@English{\l@english}
\def\l@eng{\l@english}
\def\l@EN{\l@english}
\def\l@de{\l@ngerman}
\def\l@deu{\l@ngerman}
\def\l@Deutsch{\l@ngerman}
\def\l@DE{\l@ngerman}
\def\l@ger{\l@ngerman}
\def\l@German{\l@ngerman}
\makeatother

\usepackage[iso,english,ngerman]{isodate}\isodash{--} %
\usepackage{datetime} %

\date{\today}

\newcommand{\defaultauthor}{Stefan Tauner}

\newcommand{\defaultemailraw}{stauner@ecs.tuwien.ac.at}
\newcommand{\defaultemail}{\mbox{\href{mailto:\defaultemailraw}{\defaultemailraw}}}

\author{%
  \defaultauthor\\%
  {\smaller\defaultemail}%
}

\usepackage{lipsum}
\usepackage{textcomp}
\usepackage{graphicx}
\usepackage{letltxmacro}
\LetLtxMacro\oldincludegraphics\includegraphics
\renewcommand{\includegraphics}[2][]{%
  \IfFileExists{#2}{%
    \oldincludegraphics[#1]{#2}%
  }{%
    \typeout{No file #2.}%
  }
}

\usepackage{titlecaps} %
\Addlcwords{a aboard about above across after against along amid among an and anti around as aside at atop before behind below beneath beside besides between beyond but by circa concerning considering despite down during except excepting excluding following for from given in inside into lest like mid midst minus near next nor of off on onto opposite or out outside over pace past per plus pro qua regarding round sans save since so than the through thru till times to toward towards under underneath unlike until unto up upon versus vice with within without worth yet}

\usepackage{xcolor}

\definecolor{STverydarkgrey}{HTML}{3C3C3C}
\definecolor{STdarkgrey}{HTML}{727272}
\definecolor{STgrey}{HTML}{CACACA}
\definecolor{STlightgrey}{HTML}{E6E6E6}

\definecolor{STdarkgreen}{HTML}{1C820E}
\definecolor{STgreen}{HTML}{89D57F}
\definecolor{STdarkred}{HTML}{A53A1D}
\definecolor{STred}{HTML}{EB957E}
\definecolor{STdarkblue}{HTML}{0047A0}
\definecolor{STblue}{HTML}{72A6E7}
\definecolor{STlightblue}{HTML}{B9D8FF}
\definecolor{STviolet}{HTML}{BB99CF}
\definecolor{STturquoise}{HTML}{59C7B2}
\definecolor{STdarkorange}{HTML}{C87B0F}
\definecolor{STorange}{HTML}{FFC97D}
\definecolor{STyellow}{HTML}{FFFF7D}

\usepackage[binary-units]{siunitx} %

\usepackage{booktabs}
\usepackage{multirow}
\usepackage{array}

\usepackage{algorithm}
\usepackage{algorithmicx}
\usepackage{algpseudocode}
\usepackage{listings}
\lccode`\/=0 %

\usepackage[obeyFinal]{todonotes}
\presetkeys{todonotes}{inline}{}
\LetLtxMacro{\oldtodo}{\todo}

\ifdefmacro{\hidetodos}{%
  \renewcommand\todo[2][]{}
}{
  \renewcommand\todo[2][]{\vspace{0.5ex}\oldtodo[nolist, color={yellow!15}, #1]{#2}}
}

\usepackage{soul}
\definecolor[named]{hlfixcolor}{cmyk}{0,0,0,0.1} 
\sethlcolor{hlfixcolor}
\usepackage{mychangebar2018}
\usepackage{mparhack}
\ifdefmacro{\hidetodos}{%
  \newenvironment{hlfixl}[1]{}{}
}{
  
}

\ifdefmacro{\hidehighlights}{%

}{
  
}

\usepackage[sort&compress,numbers]{natbib} %

\usepackage[nomain,nopostdot,acronym,shortcuts,nonumberlist]{glossaries-prefix}
\usepackage{glossary-mcols}
\newglossarystyle{mymcolindex}{%
  \setglossarystyle{mcolindex} %
  \renewcommand*{\glossentry}[2]{%
     \item\glsentryitem{##1}\glstreenamefmt{\glstarget{##1}{\glossentryname{##1}}}%
     \ifglshassymbol{##1}{\space(\glossentrysymbol{##1})}{}%
     \glstreepredesc \titlecap{\glossentrydesc{##1}}\glspostdescription\space ##2%
  }%
  \renewcommand{\subglossentry}[3]{%
    \ifcase##1\relax
      \item
    \or
      \subitem
      \glssubentryitem{##2}%
    \else
      \subsubitem
    \fi
    \glstreenamefmt{\glstarget{##2}{\glossentryname{##2}}}%
    \ifglshassymbol{##2}{\space(\glossentrysymbol{##2})}{}%
    \glstreechildpredesc\titlecap{\glossentrydesc{##2}}\glspostdescription\space ##3%
  }%
}
\setglossarystyle{mymcolindex} %

\newcommand{\acro}[2]{\newacronym{#1}{#1}{#2}}
\newcommand{\acroa}[2]{\newacronym[prefix={a\ },]{#1}{#1}{#2}} %
\newcommand{\acroan}[2]{\newacronym[prefixfirst={an\ },prefix={an\ },prefixfirstplural={an\ },prefixplural={an\ },]{#1}{#1}{#2}} %
\newcommand{\acroanl}[2]{\newacronym[prefixfirst={a\ },prefix={an\ },]{#1}{#1}{#2}} %

\makeglossaries
\glsdisablehyper %
\acroan{ABFT}{algorithm based fault tolerance}
\acroan{ABI}{application binary interface}
\acroan{ACET}{average-case execution time}
\acroan{ACS}{Authenticated Call Stack} %
\acroan{A/D}{Accessed/Dirty} %
\acro{ADI}{Application Data Integrity} %
\acroan{AE}{authenticated encryption}
\acroan{AEE}{application execution environment} %
\acroan{AEP}{Asynchronous Exit Pointer} %
\acroan{AET}{Architecture Event Trace} %
\acroan{AEX}{Asynchronous Enclave Exit} %
\acroan{AFI}{\ac{AXI} FIFO Interface} %
\acroan{AGU}{address-generation unit}
\acroan{ALU}{arithmetic logic unit}
\acro{AMAT}{average memory access time}
\acro{AMS}{analog-mixed-signal}
\acro{APB}{Advanced Peripheral Bus}
\acro{APE}{Authenticated Permutation-Based Encryption}
\acro{ARM}{Acorn \acs{RISC} Machine}\glsunset{ARM}
\acroan{ASIC}{application-specific integrated circuit}\glsunset{ASIC}
\acroan{ASLR}{address space layout randomization}
\acro{ATG}{automated test generation}
\acroan{AU}{arbitrary unit}
\acroan{AVF}{architectural vulnerability factor}
\acro{AXI}{Advanced eXtensible Interface}

\acro{BAH}{branch address hashing} %
\acro{BB}{basic block}
\acro{BCET}{best-case execution time}
\acro{BE}{best-effort}
\acro{BFT}{byzantine fault tolerance}
\acro{BGI}{byte granularity isolation}
\acro{BHT}{branch history table}
\acro{BIST}{built-in self test}
\acro{BLAS}{basic linear algebra subprograms}
\acro{BOOM}{Berkeley Out-of-Order Machine} %
\acro{BOV}{buffer overflow vulnerability}
\acro{BPSG}{borophosphosilicate glass}
\acro{BR}{Branch Regulation}
\acro{BRAM}{block \glsentrylong{RAM}}\glsunset{BRAM}
\acro{BTB}{branch target buffer}
\acro{BTI}{bias temperature instability} %
\acro{BTS}{Branch Trace Store} %

\newacronym[prefixfirst={a\ },prefix={an\ },]{CPSep}{CPS}{code pointer separation} %
\newacronym[prefixfirst={a\ },prefix={an\ },]{CPSys}{CPS}{cyber-physical system}
\acro{CA}{computer architecture}
\acro{CAM}{content-addressable memory}
\acro{CARS}{Collaborative Autonomous \& Resilient Systems} %
\acro{CCFI}{cryptographic \glsentrylong{CFI}}
\acro{CCP}{cache coherence protocol}
\acro{CDC}{clock domain crossing}
\acro{CED}{concurrent error detection}
\acro{CEMA}{correlation electromagnetic analysis} %
\acro{CEP}{complex event processing}
\acro{CET}{Control-flow Enforcement Technology} %
\acro{CFC}{control flow checking}
\acro{CFE}{control flow error}
\acro{CFG}{control flow graph}
\acro{CFI}{control flow integrity}
\acro{CHERI}{Capability Hardware Enhanced \acs{RISC} Instructions} %
\acro{CISC}{complex instruction set computer}\glsunset{CISC}
\acro{CLB}{configurable logic block}
\acro{CMP}{chip multiprocessor}
\acro{COP}{call-oriented programming}
\acro{COTS}{commercial off-the-shelf}\glsunset{COTS}
\acro{CPI}{code pointer integrity}
\acro{CPU}{central processing unit}\glsunset{CPU}
\acro{CRA}{code-reuse attack}
\acro{CRC}{cyclic redundancy check}\glsunset{CRC}
\acro{CRT}{constrained-random test}
\acro{CSM}{continuous signature monitoring} %
\acro{CSR}{Control and Status Register} %
\acro{CTE}{constant time execution}
\acro{CUT}{circuit under test}

\acro{DAC}{digital-analog converter}\glsunset{DAC}
\acro{DBT}{dynamic binary translation}
\acro{DCM}{digital clock manager} %
\acro{DC-RES}{Doctoral College Resilient Embedded Systems}
\acro{DCS}{distributed control systems} %
\acro{DECTED}{double-error correction, triple-error detection}
\acro{DEMA}{differential electromagnetic analysis} %
\acro{DEP}{data execution prevention} %
\acro{DFA}{differential fault attack}
\acro{DFC}{data flow checking}
\acro{DFI}{data flow integrity}
\acro{DIFT}{dynamic information flow tracking}
\acro{DLP}{data leak prevention}
\acro{DMB}{Data Memory Barrier} %
\acro{DMR}{dual modular redundancy}
\acro{DOP}{data-oriented programming}
\acro{DPA}{differential power analysis} %
\acro{DP}{double precision}
\acro{DPR}{dynamic partial reconfiguration}
\acro{DRAM}{dynamic \glsentrylong{RAM}}\glsunset{DRAM}
\acro{DSB}{Data Synchronization Barrier} %
\acro{DSP}{digital signal processors}
\acro{DTLB}{data \ac{TLB}}
\acro{DTM}{Debug Transport Module} %
\acro{DTMR}{distributed \ac{TMR}}
\acro{DUE}{detected unrecoverable error}
\acro{DUT}{design under test}
\acro{DVFS}{dynamic voltage and frequency scaling}
\acro{DWC}{duplicate with compare}

\acroan{ECC}{error-correcting code}
\acroan{ECSS}{European Cooperation for Space Standardization} %
\acroan{ECU}{electronic control unit}
\acro{EDA}{electronic design automation}
\acroan{EDC}{error-detecting code}
\acroan{EDIF}{electronic design interchange format}\glsunset{EDIF}
\acroan{EDM}{error detection mechanism}
\acroan{EEI}{execution environment interface}
\acroan{ELF}{early-life failure}
\acroan{EPC}{exception \ac{PC}} %
\acroan{EPC-SGX}{enclave page cache} %
\acroan{EPT}{extended page table} %
\acroan{ESA}{European Space Agency}\glsunset{ESA}
\acro{EX}{execution}
\acro{EXCEC}{EXtensive CFI Enforcement Concept}

\acro{FC}{fabric controller} %
\acro{FDIR}{fault detection, isolation and recovery}
\newacronym[plural=FFs,longplural=flip-flops]{FF}{FF}{flip-flop}
\acroanl{FIC}{Fault Injection Controller} %
\acroanl{FI}{fault injection}
\acroanl{FIJIEE}{\glsentrylong{FIJI} Execution Engine} %
\acroanl{FIJI}{Fault InJection Instrumenter}\glsunset{FIJI}
\acroanl{FIT}{failures in time}
\acroanl{FIU}{Fault Injection Unit} %
\acroanl{FPGA}{field-programmable gate array}\glsunset{FPGA}
\acroanl{FPU}{floating-point unit}\glsunset{FPU}
\acro{FS}{functional safety}
\acroanl{FSM}{finite state machine}
\acro{FSP}{fault-space pruning}
\acro{FT}{fault tolerance}
\acro{FU}{functional unit}

\acro{GA}{genetic algorithm} %
\acro{GALS}{globally asynchronous, locally synchronous}
\acro{GE}{gate equivalent}
\acro{GPA}{guest physical address}
\acro{GPSA}{generalized path signature analysis} %
\acro{GPU}{graphics processing unit}
\acro{GVA}{guest virtual address}

\acroa{Hart}{hardware thread}
\acroanl{HBI}{hypervisor binary interface} %
\acroanl{HCD}{Hot Carrier Degradation} %
\acro{HCFI}{Hardware-enforced Control-Flow Integrity}
\acro{HCI}{hot-carrier injection} %
\acroanl{HDL}{hardware description language}\glsunset{HDL}
\acroanl{HEE}{hypervisor execution environment}
\acro{HIL}{hardware-in-the-loop}
\acroanl{HLS}{high-level synthesis}
\acroanl{HPC}{high-performance computing}
\acroanl{HRTT}{hard real-time thread}
\acroanl{HSA}{heterogeneous system architecture}
\acroanl{HTM}{hardware transactional memory}
\acroanl{HVM}{high volume manufacturing}
\acroanl{HW}{hardware}\glsunset{HW}
\acro{HWPE}{Hardware Processing Extension} %

\acro{IACA}{Intel Architecture Code Analyzer}
\acroan{IA}{information assurance}
\acroan{IAT}{Import Address Table} %
\acro{IBT}{Indirect Branch Tracking} %
\acro{ICAP}{Internal Configuration Access Port} %
\acroan{IC}{integrated circuit}\glsunset{IC}
\acroan{ICRI}{Intel Collaborative Research Institute}
\acroan{ICS}{industrial control system}
\acro{ID}{intstruction decode}
\acroan{IDT}{interrupt descriptor table}
\acroan{IE}{Innovation Engine} %
\acro{IF}{intstruction fetch}
\acro{IFS}{information flow safety}
\acro{IFT}{information flow tracking}
\acroan{ILP}{integer linear programming}
\acroan{ILT}{Import Lookup Table} %
\acroan{IOB}{I/O block}
\acroan{iO}{in-order}
\acroan{IO}{input/output}\glsunset{IO}
\acroan{IoT}{Internet of things}\glsunset{IoT}
\acroan{IPI}{inter-processor interrupt}
\acroan{IP}{intellectual property}\glsunset{IP}
\acroan{IR}{intermediate representation}
\acroan{IRM}{inlined reference monitors}
\acroan{IRP}{I/O request packet} %
\acroan{IRQ}{interrupt request}
\acroan{ISA}{instruction set architecture}
\acroan{ISB}{Instruction Synchronization Barrier} %
\acroan{ISR}{interrupt service routine}
\acroan{IT}{intrusion tolerance}
\acro{ITLB}{instruction \ac{TLB}}

\acro{JOP}{jump-oriented programming}

\acroanl{LBR}{Last Branch Record} %
\acro{LET}{linear energy transfer} %
\acro{LFB}{line-fill buffer} %
\acroanl{LFSR}{linear feedback shift register}\glsunset{LFSR}
\acroanl{LHS}{left-hand side}\glsunset{LHS}
\acroanl{LoC}{lines of code}
\acro{LSB}{least significant bit}\glsunset{LSB}
\acroanl{LSR}{Label State Register} %
\acroanl{LSS}{Label State Stack} %
\acroanl{LSTM}{long short-term memory} %
\acroanl{LSU}{load-store unit}
\acro{LTL}{Linear Temporal Logic}
\acro{LTO}{link-time optimization}
\newacronym[plural=LUTs,longplural=look-up tables]{LUT}{LUT}{look-up table}

\acroanl{MAC}{message authentication code}
\acroanl{MBTag}{multi-bit tag}
\acroanl{MBU}{multiple-bit upset}
\acroanl{MCS}{mixed criticality system}
\acroanl{MCU}{multi-cell upset}
\acroanl{MER}{module-based error recovery}
\acro{MH}{metaheuristic}
\acro{MIL}{model-in-the-loop}
\acroanl{MI}{memory introspection}
\acroanl{MMU}{memory management unit}
\acroanl{MoC}{model of computation}
\acroanl{MPK}{Memory Protection Key} %
\acroanl{MP}{multiprocessing}
\acroanl{MPPA}{multi-purpose processor array}
\acroanl{MPU}{memory protection unit}
\acroanl{MPX}{Memory Protection Extensions} %
\acro{MSB}{most significant bit}\glsunset{MSB}
\acroanl{MTBU}{mean time between upsets}
\acro{MTE}{Memory Tagging Extension} %
\acro{MTL}{Metric Temporal Logic}
\acroanl{MT}{Merkle tree} %
\newacronym[plural=MUXes,longplural=multiplexers]{MUX}{MUX}{multiplexer}

\acroanl{NDP}{near-data processing}
\acroanl{NI}{network interface}
\acroan{NMR}{n-modular redundancy}
\acroanl{NPT}{nested page table}
\acroanl{NX}{no-execute} %

\acroan{OCD}{on-chip debugger}
\acroan{OoO}{out-of-order}
\acroan{OR1K}{OpenRISC 1000}
\newacronym[plural={OSes},prefix={an\ },]{OS}{OS}{operating system}

\acro{PAC}{pointer authentication code} %
\acro{PAN}{privileged access never} %
\acro{PA}{pointer authentication}
\acro{PCIe}{PCI Express}
\acroa{PC}{program counter}
\acroa{PDF}{path-delay fault}
\acroa{PersComp}{personal computer}\glsunset{PersComp}
\acroa{PF}{page fault}
\acro{PIL}{processor-in-the-loop}
\acroa{PIP}{programmable interconnect point}
\acro{PKEYS}{Memory Protection Keys} %
\acroa{PKU}{Memory Protection Keys for Userspace} %
\acro{PL}{programmable logic}
\acroa{PLL}{phase-locked loop}\glsunset{PLL}
\acro{PLR}{process-level redundancy}
\acroa{PMA}{physical memory protection} %
\acro{PMCA}{programmable many-core accelerator}
\acro{PMP}{physical memory protection} %
\acro{PoC}{proof of concept} \glsunset{PoC}
\acroa{PPN}{physical page number}
\acroa{PRNG}{pseudo random number generator}
\acro{PTI}{page table isolation}
\acroa{PTW}{page table walker}
\acroa{PUF}{physically unclonable function}
\acro{PVT}{process, voltage and temperature}
\acro{PXN}{privileged execute never} %

\acro{QDI}{quasi-delay-insensitive}

\acro{RAM}{random access memory}\glsunset{RAM}
\acro{RAR}{return address repository}
\acro{RAS}{return address stack}
\acroanl{RAT}{register alias table}
\acroanl{RC}{reconfiguration controller}
\acroanl{RCU}{read-copy-update}
\acroanl{REV}{runtime execution validator}
\acroanl{RF}{register file}
\acroanl{RISC}{reduced instruction set computer}%
\acroanl{RMT}{redundant multithreading}
\acroanl{RNG}{random number generator}
\acroa{RoCC}{Rocket Custom Coprocessor}
\acroanl{ROP}{return-oriented programming}
\acroanl{RSE}{run-time scope enforcement}
\acroanl{RTL}{register-transfer level}\glsunset{RTL}
\acroanl{RTS}{real-time system}
\acroanl{RVI}{rapid virtualization indexing} %
\acroanl{RV}{runtime verification}
\acroanl{RVTSO}{RISC-V Total Store Ordering}
\acroanl{RVWMO}{RISC-V Weak Memory Ordering}

\newacronym[prefixfirst={a\ },prefix={an\ },]{StatFI}{SFI}{statistical fault injection}
\newacronym[prefixfirst={a\ },prefix={an\ },]{SwFI}{SFI}{software fault isolation}
\acroanl{SBI}{supervisor binary interface} %
\acro{SBO}{surrogate-based optimization} %
\acroanl{SBSE}{search-based software engineering}
\acro{SBST}{search-based software testing}
\acroanl{SBTag}{single-bit tag}
\acroanl{SBT}{search-based testing}
\acro{SCADA}{supervisory control and data acquisition} %
\acroanl{SCA}{side-channel attack}
\acro{SCFP}{Sponge-based Control Flow Protection}
\acroanl{SCS}{shadow call stack} %
\acroanl{SDC}{silent data corruption}
\acroanl{SDM}{spatial division multiplexing}
\acroanl{SEA}{speculative execution attack} %
\acroanl{SEB}{single-event burnout}
\acro{SECDED}{single error correction, double error detection}
\acroanl{SEE}{single-event effect}
\acroanl{SEFI}{single event functional interrupt}
\acroanl{SEGR}{single-event gate rupture}
\acroanl{SEL}{single-event latch-up}
\acroanl{SEM}{soft error mitigation}
\acroanl{SER}{soft error rate}
\acro{SES}{soft error sensitivity}
\acroanl{SET}{single-event transient}
\acroanl{SEU}{single-event upset}
\acro{SEV}{Secure Encrypted Virtualization} %
\acroanl{SGX}{Software Guard Extensions} %
\acro{SIL}{software-in-the-loop}
\acroanl{SIS}{signatured instruction stream}
\acroanl{SLAT}{second level address translation} %
\acroanl{SLR}{systematic literature review}
\acroanl{SMAP}{supervisor mode access protection} %
\acroanl{SMEP}{supervisor mode execution protection} %
\acro{SME}{Secure Memory Encryption} %
\acro{SMR}{state machine replication}
\acroanl{SM}{switch matrix}
\acroanl{SMT}{simultaneous multithreading}
\acroanl{SoK}{State of Knowledge}\glsunset{SoK}
\acroanl{SotA}{state of the art}
\acro{SPARC}{Scalable Processor Architecture}\glsunset{SPARC} %
\acroanl{SPA}{simple power analysis} %
\acroanl{SPN}{Standard Pattern Name} %
\acroanl{SPP}{sub-page permission}
\acroanl{SPPTP}{sub-page permission table pointer}
\acroanl{SPPT}{sub-page permission table}
\acroanl{SRAM}{static \glsentrylong{RAM}}\glsunset{SRAM}
\acro{SRAS}{secure \ac{RAS}}
\acroanl{SRS}{Storage Region Stack} %
\acroanl{SRTR}{redundantly threaded processors with recovery}
\acroanl{SRTT}{soft real-time thread}
\acroanl{SSBS}{Speculative Store Bypass Safe} %
\acro{SSDT}{secondary system description table} %
\acroanl{SSP}{Stack Smashing Protector} %
\acroanl{STA}{static timing analysis}
\acroanl{STL}{Signal Temporal Logic}
\acro{SUS}{system under test}
\acro{SVA}{SystemVerilog Assertions}
\acro{SVF}{side channel vulnerability factor}

\acro{TBI}{Top-Byte Ignore} %
\acro{TCB}{trusted computing base}
\acro{TCR}{Tag Check Register} %
\acroa{TDDB}{time-dependent dielectric breakdown} %
\acroa{TDL}{tapped delay line} %
\acro{TDM}{time-division multiplexing}
\acroa{TEE}{trusted execution environment}
\acro{TE}{trusted execution}
\acroa{TID}{total ionizing dose} %
\acro{TI}{threshold implementation} %
\acro{TLB}{translation lookaside buffer}
\acro{TLP}{thread-level parallelism}
\acro{TMR}{triple modular redundancy}
\acroa{TM}{transactional memory}
\acro{TPR}{Tag Propagation Register} %
\acroa{TRNG}{true random number generator}
\acro{TSM}{trusted software module}
\acro{TSN}{time-sensitive networking}
\acro{TSV}{through-silicon via}
\acro{TSX}{Transactional Synchronization Extensions} %
\acroa{TVLA}{test vector leakage assessment} %

\acroa{UAF}{use after free}
\acroa{UART}{universal asynchronous receiver/transmitter}\glsunset{UART}
\acroa{ULFM}{user level failure mitigation}
\acroa{UMIP}{user mode instruction prevention}
\acroa{UUT}{unit under test}

\acro{VGA}{video graphics array}\glsunset{VGA}
\acro{VHDL}{{VHSIC} Hardware Description Language}\glsunset{VHDL}
\acro{VHSIC}{Very High Speed Integrated Circuits}\glsunset{VHSIC}
\acro{VLIW}{very large instruction word}
\acro{VMI}{virtual machine introspection}
\acroa{VMM}{virtual machine monitor}
\acro{VM}{Verilog Mapping} %
\acroa{VPN}{virtual page number}
\acro{VQM}{Verilog Quartus Mapping} %

\acro{WBC}{white-box cryptography}
\acro{WB}{write back}
\acro{WCET}{Worst-Case Execution Time}
\acro{WED}{weighted edit distance}
\acro{WIP}{work in progress}\glsunset{WIP}
\newacronym{WLOG}{w.l.o.g.}{without loss of generality}

\acroanl{XD}{execute disable} %
\acroanl{XN}{execute never} %

\newacronym[longplural={networks on chips},prefixfirst={a\ },prefix={an\ },]{NoC}{NoC}{network on chip}
\newacronym[longplural={page table entries},prefixfirst={a\ },prefix={an\ },]{PTE}{PTE}{page table entry}
\newacronym[longplural={scratch-pad memories,prefixfirst={a\ },prefix={an\ },}]{SPM}{SPM}{scratch-pad memory}
\newacronym[longplural={single-event multiple upsets},shortplural={SEMUs}]{SEMU}{SEMUs}{single-event multiple upsets}
\newacronym[longplural={systems on chips,prefixfirst={a\ },prefix={an\ },}]{SoC}{SoC}{system on chip}\glsunset{SoC}
\newacronym[longplural={tightly-coupled data memories,prefixfirst={a\ },prefix={a\ },}]{TCDM}{TCDM}{tightly-coupled data memory}
\newcounter{acronum}
\makeatletter
  \newglossarystyle{countentries}{%
    \setcounter{acronum}{0}%
    \let\glossaryheader\@empty%
    \let\glsgroupheading\@gobble%
    \let\glsgrouptitle\@gobble%
    \let\glsnavhypertarget\@gobbletwo%
    \let\glsnavigation\@empty%
    \let\glsgroupskip\@empty%
    \let\glsentryitem\@gobble%
    \let\glsentrycounterlabel\@empty%
    \let\glstarget\@gobbletwo%
    \let\glossaryentrynumbers\@gobble%
    \let\subglossentry\@gobblethree%
    \let\glssubentryitem\@gobble%
    \let\glssubentrycounterlabel\@empty%
    \let\currentglossary\@empty%
    \renewcommand\glossarysection[2][]{}%
    \let\glossarypreamble\@empty%
    \let\glossarypostamble\@empty%
    \let\glsresetentrylist\@empty%
    \renewcommand\glossentry[2]{\stepcounter{acronum}}%
  }
  \AtBeginDocument{\printacronyms[style=countentries]}

\usepackage{hyperref}
\hypersetup{
  pdfstartview={FitV},
  pdfpagelayout=SinglePage,
  pdfpagemode=UseOutlines,
  bookmarksopen=true,
  bookmarksopenlevel=3,
  colorlinks=false, %
  allbordercolors=black, %
  pdfborderstyle={/S/U/W 0.1}, %
  pdflinkmargin={1.5pt},
  pdfauthor={\defaultauthor},
}

\addto\extrasenglish{

}

\makeatletter
\LetLtxMacro{\oldfootnotetext}{\@footnotetext}
\renewcommand{\@footnotetext}[1]{\oldfootnotetext{\hypersetup{hidelinks}#1}}
\LetLtxMacro{\oldmpfootnotetext}{\@mpfootnotetext}
\renewcommand{\@mpfootnotetext}[1]{\oldmpfootnotetext{\hypersetup{hidelinks}#1}}
\makeatother

\usepackage[nameinlink]{cleveref}
\crefname{subsection}{subsection}{subsections}

\newif\ifshowbib
\showbibfalse
\def\citation#1{\global\showbibtrue}
\let\oldbibliography\bibliography
\renewcommand{\bibliography}[1]{\ifshowbib\oldbibliography{#1}\fi}

\makeatletter
\if@twocolumn
  
\else
  
\fi
\makeatother

\usepackage{xcolor}
\definecolor{col1}{rgb}{1.0,0.01,0.24}      %
\definecolor{col2}{rgb}{0,0,1.0}            %
\definecolor{col3}{rgb}{1.0,0.49,0}         %
\definecolor{col4}{rgb}{0.01,0.75,0.24}     %
\definecolor{col5}{rgb}{0.89,0.82,0.04}     %
\definecolor{col6}{rgb}{0.28,0.24,0.2}      %

\usepackage{adjustbox}
\newcolumntype{R}[2]{%
    >{\adjustbox{angle=#1,lap=\width-(#2)}\bgroup}%
    l%
    <{\egroup}%
}
\newcommand*\rot{\multicolumn{1}{R{45}{1em}}}

\newcommand*\emptycirc{\tikz\draw (0,0) circle (.75ex);} 

\newcommand*\fullcirc{\tikz\fill (0,0) circle (.75ex);} 

\usepackage{colortbl}

\usepackage{pgfplots}
\pgfplotsset{compat=1.16}
\usepackage{tikz}
\usetikzlibrary{shapes,arrows,backgrounds,fit,positioning, pgfplots.statistics, pgfplots.colorbrewer} 

\usepackage{marvosym}

\usepackage{stfloats} %

\hypersetup{
  pdfauthor={},
}

\usepackage{multibib} \usepackage{fancyhdr}

\def\BibTeX{{\rm B\kern-.05em{\sc i\kern-.025em b}\kern-.08em
    T\kern-.1667em\lower.7ex\hbox{E}\kern-.125emX}}

\author{%
  \ifdefmacro{\review}{%
    \IEEEauthorblockN{Anonymized For Review}
    \IEEEauthorblockA{%
      \textit{Department} \\
      \textit{Affiliation}%
      \\
      \mbox{\href{example@example.com}{example@example.com}}%
    }

    \vphantom{\thanks{}}
  }{%
    \IEEEauthorblockN{Mario Telesklav}
    \IEEEauthorblockA{%
      \textit{Embedded Computing Systems Group} \\
      \textit{TU Wien}%
      \\
      \mbox{\href{mailto:mario.telesklav@ecs.tuwien.ac.at}{mario.telesklav@ecs.tuwien.ac.at}}%
    }
    \and
    \IEEEauthorblockN{Stefan Tauner}
    \IEEEauthorblockA{%
      \textit{Embedded Computing Systems Group} \\
      \textit{TU Wien}%
      \\
      \mbox{\href{mailto:\defaultemailraw}{\defaultemailraw}}%
    }%
    \thanks{The authors contributed equally to this work.}
  }%
}
\IEEEoverridecommandlockouts %
 
\fancypagestyle{firstpage}{
  \fancyhf{}
}  

\title{%
  Comparative Analysis and Enhancement of CFG-based Hardware-Assisted CFI Schemes
}

\newcites{Links}{Links}

\makeatletter%
\begin{document}
\maketitle
\pagestyle{plain}

\begin{abstract}
Subverting the flow of instructions (e.g., by use of \aclp{CRA}) still poses a serious threat to the security of today's systems.
Various \ac{CFI} schemes have been proposed as a powerful technique to detect and mitigate such attacks.
In recent years, many hardware-assisted implementations of \ac{CFI} enforcement based on \acp{CFG} have been presented by academia.
Such approaches check whether control flow transfers follow the intended \ac{CFG} by limiting the valid target addresses.
However, these papers all target different platforms and were evaluated with different sets of benchmark applications, which makes quantitative comparisons hardly possible.

For this paper, we have implemented multiple promising \ac{CFG}-based \ac{CFI} schemes on a common platform comprising a RISC-V \ac{SoC} within \pgls{FPGA}.
By porting almost 40 benchmark applications to this system we can present a meaningful comparison of the various techniques in terms of run-time performance, hardware utilization, and binary size.
In addition, we present an enhanced \ac{CFI} approach that is inspired by what we consider the best concepts and ideas of previously proposed mechanisms.
We have made this approach more practical and feature-complete by tackling some problems largely ignored previously.
We show with this fine-grained scheme that \ac{CFI} can be achieved with even less overheads than previously demonstrated.
 \end{abstract}

\section{Introduction}
\label{sec:intro}

The eventual goal of any attacker is to manipulate the execution flow of a program, e.g., to exfiltrate data by overcoming restrictions or to execute malicious code.
To that end, input data is manipulated to exploit the abstract machines that programming languages/environments and the underlying hardware provide.
Prominent examples are the notorious class of stack buffer overflows~\cite{levySmashingStackFunProfit1996} and format string vulnerabilities~\cite{scutExploitingFormatStringVulnerabilities2001}.
These approaches take advantage of the fact that targets of indirect control flow transfers like return addresses of functions are often stored on the stack where they can easily be overwritten if certain faulty constructs end up in programs.

In earlier days one could not only change the control flow that way but even directly supply malicious code to be executed from the stack.
However, in the last two decades defenses against these and more advanced threats have been improved.
Unfortunately, there is still no silver bullet in sight:
Static and dynamic bug finding tools~\cite{songSoKSanitizingSecurity2019} have become able to warn programmers of the exploitable bugs before they creep into the code base (but they have to be utilized);
\Acp{CPU} (often even those in embedded systems) can enforce non-executability of certain address ranges like the stack (if set up correctly~\cite{abbasiChallengesDesigningExploitMitigations2019});
\Ac{ASLR} makes it harder (but not impossible~\cite{goktasPositionindependentCodeReuseEffectiveness2018}) to derive the addresses necessary to mount return-to-libc attacks or exploit use-after-free vulnerabilities;
sandboxing and splitting applications into multiple (micro)services simplifies logic locally (but increases overall complexity)~\cite{randalIdealRealRevisitingHistory2020}.

\subsection{Code-Reuse Attacks}

While the improved defense mechanisms have triggered an arms race with attackers that made possible attacks very sophisticated~\cite{samuelgrossProjectZeroJITSploitationIII2020}, the basic ideas have remained the same:
\begin{enumerate*}
  \item establish a write primitive to manipulate the stack or heap,
  \item determine a suitable address to overwrite,
  \item overwrite data there to branch off into an unintended flow of instructions.
\end{enumerate*}

Due to the mitigations typically employed in modern systems it is usually not possible to simply inject new instructions to be executed.
Therefore, adversaries rely on the existing code to chain together snippets called \textit{gadgets} in \acp{CRA}.
This can, for example, be accomplished by jumping to a short code sequence ending in a \texttt{ret} whose target address was manipulated in \ac{ROP}~\cite{shachamGeometryInnocentFleshBone2007}, or exploit similar semantics of indirect jumps in \ac{JOP}~\cite{checkowayReturnorientedProgrammingReturns2010}.
In either case the attacker can produce Turing-complete arrangements by completely mangling the sequence of instructions intended by the programmer.

\subsection{Control Flow Integrity}

The intended flow is captured by the \acf{CFG} of a program consisting of \acp{BB} connected by control transfer instructions.
Depending on the \ac{ISA} the latter consist of instructions to call and return from functions, direct and indirect jumps etc.
The connections in the \ac{CFG} spanned by returns are called \textit{backward} edges while all other control transfers are \textit{forward}.

To counter attacks that try to break out of the intended \ac{CFG}, various forms of \acf{CFI} enforcement ideas have been presented.
Interestingly, the earliest concepts to guarantee \ac{CFI} have not been proposed for security but safety reasons.
In the 1990s Wilken and Shen developed \acf{CSM} as a countermeasure against transient faults in the control logic of processors~\cite{wilkenContinuousSignatureMonitoringLowcost1990}.

The groundwork for nowadays security-centered \ac{CFI} was laid in 2005 by Abadi et al.\ with their original \ac{CFI} proposal~\cite{abadi2009control}.
The attack model comprises an adversary with full control of the entire data memory (while code is thought to be read-only).
Under these assumptions, direct calls and jumps do not pose a threat because their target cannot be changed.
The main idea is to enforce that every indirect control flow transfer only targets one of its allowed destinations.
The authors suggest to guarantee this by checking labels inserted at the respective target \acp{BB}.
Every indirect call, jump, and also return has to verify that their inherent label from the originating \ac{BB} matches the target.
This basic approach is independent of the \ac{ISA} and can be implemented in software only or by extending the hardware to handle the checks directly.

In this work we focus on \ac{CFI} implementations that rely on hardware support.
For a review of software-only approaches we refer to the survey by Burow et al.~\cite{burow2017control}.

Even when \ac{CFI} is able to enforce a fully-precise \ac{CFG} it is not very effective on its own.
The problem stems from the fact that without a run-time state it is impossible to determine securely if a return address is matching the correct caller or if it was manipulated to follow another allowed edge of the \ac{CFG}.
For that reason backward edges have to be protected by additional mechanisms, e.g. by a shadow stack, to remain effective~\cite{carliniControlflowBendingEffectivenessControlflow2015}.

A \textit{shadow stack} (or return address stack) is an extra protected memory for redundantly keeping track of return addresses.
Its basic concept was already proposed in the early 2000s as \ac{RAR}%
~\cite{chiueh2001rad}.
In hardware it can be implemented without run-time overhead if the respective memory is directly attached to the \ac{CPU} pipeline.

\subsection{Instrumentation}

The process of injecting instructions into some program at certain well-defined positions is called \textit{instrumentation}.
Doing so in an already compiled binary is referred to as \textit{binary rewriting}, with a broad field of applications like emulation, optimization and profiling~\cite{wenzl2019hack}.
Using binary rewriting for instrumentation of \ac{CFI} entails that \ac{CFG} extraction is harder, especially for stripped binaries and in the absence of debug symbols.
For that reason, our implementation is based on a compiler extension that has the source code available.

\subsection{Contribution \& Outlook}

In this work we answer how well proposed schemes for hardware-assisted \ac{CFI} enforcement fare with respect to performance, hardware utilization and code size.
To that end, we have implemented selected existing \ac{CFI} protections on and ported numerous benchmark applications to a common platform allowing for a meaningful comparison.
Additionally, we improve on these schemes and present a novel ``best of'' approach that borrows the best concepts and ideas of previously proposed mechanisms.
We show with this fine-grained approach that \ac{CFI} can be achieved with even less overheads than previously demonstrated.
In the original articles evaluations have been executed on small \acp{SoC} in \acp{FPGA} due to the need of customizing the \ac{CPU} pipeline.
While none of the concepts hereinafter are necessarily limited to embedded systems we have to act under the same constraints.
However, in addition to incrementally improving the concept of hardware-assisted \ac{CFI} enforcement on small systems, we also show how to work around some practical problems that have to be solved to employ this method on larger systems.
In the next section we review the academic literature on hardware-assisted \ac{CFI} approaches.
Out of those we have selected what we think are the most diverse and promising techniques for comparison.
Their implementation details are reviewed in \autoref{sec:implementations} together with our own improved variant of \ac{CFI} enforcement.
In \autoref{sec:evaluation} we outline the test environment including benchmarks, the instrumentation process and the chosen hardware parameters.
Eventually, we present and discuss the measurement results in the categories performance, hardware utilization and code size in \autoref{sec:results} before ending with the conclusion.
\section{Related Work}
\label{sec:relwork}

We give a brief overview of existing hardware-based \ac{CFI} approaches in this section and refer the interested reader to further literature such as the extensive survey of hardware-based \ac{CFI} approaches by De Clercq and Verbauwhede in \cite{de2017survey} for more details.

\Acp{CFG} can be viewed as sequence diagrams, where some application's functions or \acp{BB} correspond to states in the diagram.
In \cite{arora2006hardware} such statically determined \ac{CFG} information is encoded into \pgls{FSM} where legit control flow transfers relate to allowed state transitions.
In the same way also intra-procedural jumps are monitored on \ac{BB} level.
Similarly, \cite{rahmatian2012hardware} focuses on system calls and encodes allowed application-specific sequences thereof in \pgls{FSM} to detect invalid executions.

SCRAP~\cite{kayaalp2013scrap} uses \pgls{FSM} for implementing a signature-based \ac{CFI} approach.
The concept monitors execution traces at runtime and applies heuristics to detect sequences of short gadgets that are often an indicator for some \ac{CRA}.
The frequency of gadget executions is also sometimes used in heuristic concepts because \acp{CRA} tend to reuse the same gadgets multiple times.

A whole other class of concepts tries to reduce the gadget space by limiting control flow transfers to \ac{BB} boundaries in some way.
\Ac{BR}~\cite{kayaalp2012efficiently} restricts control flow transfers to entries of target functions and locations within the same function.
This approach does not require \ac{CFG} information.
BB-CFI~\cite{das2016fine} went one step further and enforces that forward-edge control flow transfers only target the very beginning of certain \acp{BB}, thereby requiring a statically created \ac{CFG}.
BBB-CFI~\cite{he2017no} ensures that forward-edge control flows only target the entry of \acp{BB} in general and thereby prevents jumps to positions in the middle, which is often required for gadgets.
Similarly, Intel \ac{CET}~\cite{IntelCET} ensures that indirect jumps and calls can only target certain labels and function entries.
\Ac{CET} is more precise but requires static analysis of the control flow and additional instructions, both of which are not required in BBB-CFI.
In an abstract way, some of these approaches are variations of Abadi et al.'s original label concept and only vary granularity with the number of different labels used.

More fine-grained concepts usually require a \ac{CFG}.
HCFI~\cite{christoulakis2016hcfi} protects function calls with unique labels, close to Abadi et al.'s original proposal.
The concept presented in \cite{sullivan2016strategy} extends this approach with \textit{trampolines} in order to avoid using the same label for functions that are called from multiple callers.
Table-based concepts, where a flag indicating the validity of a certain branch target is looked up in a table upon every indirect control flow transfer, are another fine-grained but very memory-intense approach for forward-edge protection.
FIXER~\cite{de2019fixer} and \cite{arora2006hardware} represent examples thereof.

\Ac{CFI} enforcement on backward edges requires different techniques.
Most \ac{CFI} implementations use a shadow stack for that.
Examples for such proposals are HCFI~\cite{christoulakis2016hcfi}, FIXER~\cite{de2019fixer} and Intel \ac{CET}~\cite{IntelCET}, which all redundantly store the return address of function calls.
This concept was slightly modified in \cite{sullivan2016strategy}, where unique labels are stored on the stack instead of addresses.
Usually less memory is required for this variant, but instrumentation is more complex.
HAFIX~\cite{davi2015hafix} implements an alternative concept for backward-edge protection by means of an \textit{Active Set} that keeps track of currently active functions and only allows returns to functions flagged active there.
However, it has been shown that the concept is more vulnerable in comparison to shadow stacks~\cite{theodorides2017breaking}.
\section{Implementations}
\label{sec:implementations}
We have selected four of the academic fine-grained \ac{CFG}-based \ac{CFI} approaches listed above for a more detailed examination and, more importantly, implemented these concepts on a common RISC-V platform.
The criteria with which we made this selection were the precision of their \ac{CFG} enforcement as well as the variety of \ac{CFI} concepts they implement.
To widen the view we have also included a rather coarse-grained approach that is similar to a commercially available concept.
Our implementations of the presented concepts are as accurate to their originals as possible.
We have designed all of them as separate modules, which are individually attached to the core's pipeline.
Changes to the core itself only consist of extensions to its decoder for custom instructions and modifications in the pipeline controller for handling of exception signals resulting from \ac{CFI} violations.
However, we use \acp{FF} for all memory elements to enable in-cycle access for all implementations and also assume the availability of the same complete \ac{CFG} everywhere.
All memory elements are protected from deliberate software access and modification only happens implicitly during control transfers.
Whenever changes to the original concepts were required, we note them in the sections below.

\subsection{FIXER}
De et al.\ presented FIXER~\cite{de2019fixer} in 2019, which uses a shadow stack for securing backward edges. 
Forward-edge protection is implemented by means of a policy matrix, which holds flags for each pair of indirect function calls and their potential targets. 
Upon performing a call, the flag is looked up for the current combination of addresses and the call is only executed when the respective flag is set.
An address decoder translates 32-bit addresses to matrix indices for the lookup.

The authors have realized the security mechanisms as a co-processor to a RISC-V \ac{SoC}, thereby making it unnecessary to modify the actual core at all.
However, using this approach requires additional instructions for accessing pipeline registers like the current program counter.
The original FIXER concept therefore needs multiple instructions for each \ac{CFI} operation.
Given the possibility to implement the module for \ac{CFI} enforcement directly attached to the pipeline in our setup, we have removed this limitation so that only one instruction is needed.
De et al.\ reported a performance overhead of 1.5\% for backward edge protection, on average, and respectively 0.61\% for forward edges.
Area overhead was reported to be 2.9\%.

\subsection{HAFIX}
In 2015, Davi et al.\ proposed an extension to \cite{davi2014hardware} called HAFIX~\cite{davi2015hafix}. 
Their approach only covers backward-edge protection but contains two enforcement concepts, namely a shadow stack and a concept for limiting function returns to active call sites with the help of an \textit{Active Set}.
The authors argue that software-side forward edge protection is sufficiently effective.
We only consider the Active Set approach since it is their distinctive feature compared to other \ac{CFI} concepts.

Every function is marked active at the very start of its execution, and inactive when returning. 
This is done with custom instructions holding a label unique for each function that acts as an index to the Active Set.
In addition, it is enforced that functions can only return to an active function by checking the active state of the function that it is returning to. 
Simple forms of recursion are specifically handled by means of a counter measuring the recursion depth, while special cases like nested recursion are not supported.

Davi et al.\ presented HAFIX as an extension to an x86 Intel Siskiyou Peak softcore. 
The proposed concept involves an average performance overhead of around 2\%. 
2.5\% more registers and less than 1\% of additional \acp{LUT} were required.

\subsection{HCFI}
Christoulakis et al.\ presented HCFI~\cite{christoulakis2016hcfi} in 2016, which provides protection for both forward and backward edges. 
They have implemented their concept on a LEON3 SPARC softcore.
The SPARC \ac{ISA} allows for using branch delay slots to execute some of the additionally required instructions.
This is not possible on our RISC-V platform.

The authors highlight the importance of a shadow stack for \ac{CFI} enforcement of function returns and implement such with dedicated recursion handling and support for C's \texttt{setjmp}. 
Forward-edge protection is achieved by instrumenting indirect function calls and the entries of indirectly called functions with labels.
Upon calling such a function, the label set at the caller side is stored in the \ac{CFI} monitor and compared to the label in the callee. 
An internal \ac{FSM} checks that such instructions are executed back-to-back.

About 1\% run-time overhead and a hardware overhead of 2.5\% (\acp{LUT} and \acp{FF} combined) were reported. 

\subsection{Policy-Agnostic Hardware-Enhanced Control-Flow Integriy (HECFI)}
Sullivan et al.\ proposed a concept similar to HCFI in 2016 in \cite{sullivan2016strategy}, but with added support for securing indirect jumps and a very fine-grained approach for forward-edge protection. 
For easier reference we call this unnamed approach HECFI in the remainder of this work but want to note that its origins stem from HAFIX.

In HECFI, label checks for indirect calls are enriched with \textit{trampolines}.
Simple label-based approaches assign the same label to all functions, which are indirectly called from multiple places, and thus introduce some imprecision in the enforced \ac{CFG}.
Trampolines improve on that and act as an intermediate step between caller and callee where the actual target is compared to a jump table of allowed call targets.
A failed lookup in this table indicates a manipulated call and results in a \ac{CFI} exception.
With this concept, the function call is \ac{CFI}-enforced in a very fine-grained way because every indirect control flow transfer has a maximum of one assigned trampoline. 
The direct jump from the trampoline to the actual target then needs no additional \ac{CFI} protection. 
Backward edges are secured with a shadow stack.
The authors used a SPARC LEON3 processor on an \ac{FPGA} board for their evaluation. 
They reported an average run-time overhead of 1.75\%, average code size overhead of 13.5\% and area overhead of 1.78\%. 

\subsection{Intel CET}
Intel published the specification for their hardware-assisted \ac{CET} in 2016~\cite{IntelCET}.
To the best of our knowledge, the concept, which is first used in the 2020 Tiger Lake CPU generation, is the first commercially available \ac{CFI} implementation remotely comparable to the previously described \ac{CFG}-based concepts.
Intel combines a shadow stack for backward-edge protection with a coarse-grained approach for indirect branches called \ac{IBT}. 

With \ac{IBT} enabled, all targets of indirect branches are instrumented with a custom \texttt{ENDBRANCH} instruction. 
Whenever an indirect control flow transfer occurs, a state machine internal to the CPU makes sure that the transfer targets such an instruction.
This is very similar to what is described as Branch Limitation in other concepts.
Intel's coarse-grained approach eases practical problems in supporting very complex applications, where the extraction of a fine-grained \ac{CFG} might not be possible.
Presumably, it also helps integrating the concept into high-end x86 \acp{CPU}.

We have performed some basic build tests with Intel's own ICC 19.0.1, GCC 10.2 and Clang 11.0.0.
All of them are able to instrument their outputs with \texttt{ENDBRANCH} instructions to exploit this feature but it has to be enabled explicitly.
They act conservatively in the sense that \textit{any} function entry is instrumented if the compiler cannot derive with certainty that the respective function is \textit{not} called indirectly.
To allow the compilers to do so one can declare functions as \texttt{static} or enable \ac{LTO}.
In order to enable a fair comparison, we improved on this behavior and instrumented the targets of indirect control flow transfers only.

In Intel's \ac{CFI} implementation \texttt{ENDBRANCH} uses an encoding that is interpreted as a (multi-byte) \texttt{NOP} instruction by legacy hardware allowing for backwards compatibility within a single binary.
In our basic emulation of this scheme on RISC-V this is emulated by a custom 4-byte instruction.

While \ac{CET} targets a different class of processors and protection on forward edges is less fine-grained in comparison to the previously described approaches, its inclusion in our evaluation gives insights into possible gains of following a coarse-grained approach.

\subsection{\acs{EXCEC}}
We present an enhanced hardware-based \ac{CFI} scheme called \ac{EXCEC} that is based on what we consider the best concepts and ideas from the existing approaches described in the previous sections.
As we will later show in \Cref{sec:results}, \ac{EXCEC} outperforms comparable approaches in terms of run-time overhead and code size while keeping hardware complexity at similar levels.
We manage to do this while maintaining the security level of fine-grained \ac{CFI} and supporting often omitted language features that complicate implementation of the \ac{CFI} mechanism.

\ac{EXCEC} offers protection for both forward and backward facing edges.
In addition to basic \ac{CFI} enforcement on indirect function calls, we have added protection of indirect jumps, dedicated recursion handling and support for \texttt{setjmp} calls.
The custom \ac{CFI} instructions required for \ac{CFI} enforcement are shown in \autoref{tbl:cfi_instructions}.

\begin{table}[htb]
	\centering
	\begin{tabular}{ lp{5.5cm} }
		\toprule
		\textbf{Instruction} & \textbf{Description}
		\\
		\midrule
		\texttt{CFI\_CALL label} &  Announces label for indirect function call.
		Placed right before the call instruction.
		Enables label check in \ac{CFI} module.
		\\ 
		\midrule
		\texttt{CFI\_JUMP label} & Announces label for indirect jump.
		Placed right before the jump instruction.
		Enables label check in \ac{CFI} module.
		\\
		\midrule
		\texttt{CFI\_CHECK label} & Compares label with the one of a preceding \texttt{CFI\_CALL} or \texttt{CFI\_JUMP} instruction.
		Placed in the entry of indirectly called functions and at targets of indirect jumps.
		Has no effect when not enabled by a preceding indirect jump or call unless the label is \texttt{0x0} (indicating a violation).
		\\
		\midrule
		\texttt{CFI\_SETJMP index} & Stores current shadow stack pointer in \ac{CFI} module at position of \texttt{index} when no preceding \texttt{CFI\_LONGJMP} occurred.
		Otherwise, it unwinds the shadow stack to the previously stored position.
		Placed right after \texttt{setjmp} calls.
		\\
		\midrule
		\texttt{CFI\_LONGJMP} & Announces \texttt{longjmp} to the \ac{CFI} module for subsequent \texttt{CFI\_SETJMP} instruction. Placed right before calls of \texttt{longjmp}.
		\\
		\bottomrule
	\end{tabular}
	\caption{\ac{CFI} instructions for \ac{EXCEC}}
	\label{tbl:cfi_instructions}
\end{table}

\subsubsection{Forward Edges}
For protection of forward edges we use label checks between the caller and the callee, similar to what Abadi et al.\ proposed in their original \ac{CFI} paper~\cite{abadi2009control}.
Indirect function calls and jumps are annotated with custom \texttt{CFI\_CALL} or \texttt{CFI\_JUMP} instructions, which bring the \ac{CFI} module in a state where it expects a \texttt{CFI\_CHECK} with a matching label.
In addition, we borrow the concept of \textit{trampolines} from Sullivan et al.~\cite{sullivan2016strategy} in order to achieve a more fine-grained protection.

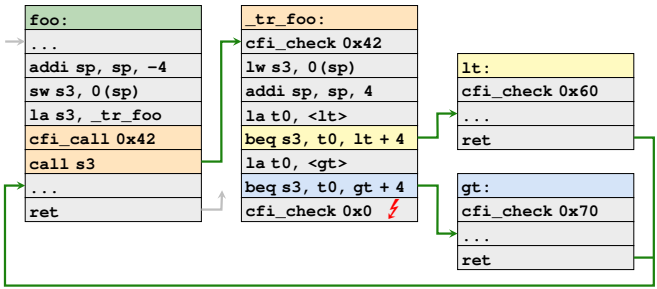
\begin{figure}[htb]
 	\centering
 	\resizebox{\linewidth}{!}{%

\tikzstyle{arrow} = [color=STdarkgreen, ultra thick,->,>=stealth]
\tikzstyle{instr}=[draw, rectangle, text height=0.4cm, minimum height=0.6cm, text width=4.2cm, minimum width=4.4cm,fill=gray!15, font=\large\bfseries\ttfamily]
		
\begin{tikzpicture}	[x=1cm, y=1cm, node distance=0,outer sep=0,inner sep=0]	
	
	\node [instr,fill=STdarkgreen!30] (foo1) {foo:};
	\node [instr] (foo2) [below = of foo1] {...} ;
	\node [instr] (foo3) [below = of foo2] {addi sp, sp, -4} ;
	\node [instr] (foo4) [below = of foo3] {sw s3, 0(sp)} ;
	\node [instr] (foo5) [below = of foo4] {la s3, \_tr\_foo} ;
	\node [instr,fill=STorange!50] (foo6) [below = of foo5] {cfi\_call 0x42} ;
	\node [instr,fill=STorange!50] (foo7) [below = of foo6] {call s3} ;
	\node [instr] (foo8) [below = of foo7] {...} ;
	\node [instr] (foo9) [below = of foo8] {ret} ;

	\node [instr,fill=STorange!50] (trfoo1) [right = of foo1, xshift=1cm] {\_tr\_foo:};
	\node [instr] (trfoo2) [below = of trfoo1] {cfi\_check 0x42} ;
	\node [instr] (trfoo3) [below = of trfoo2] {lw s3, 0(sp)} ;
	\node [instr] (trfoo4) [below = of trfoo3] {addi sp, sp, 4} ;
	\node [instr] (trfoo5) [below = of trfoo4] {la t0, <lt>} ;
	\node [instr,fill=yellow!30] (trfoo6) [below = of trfoo5] {beq s3, t0, lt + 4} ;
	\node [instr] (trfoo7) [below = of trfoo6] {la t0, <gt>} ;
	\node [instr,fill=STblue!30] (trfoo8) [below = of trfoo7] {beq s3, t0, gt + 4} ;
	\node [instr] (trfoo9) [below = of trfoo8] {cfi\_check 0x0} ;
	\node [color = red, ultra thick, right = of trfoo9, xshift=-.75cm] {\huge\textbf{\Lightning}};
	
	\node [instr,fill=yellow!30] (lt1) [right = of trfoo3, xshift=1cm] {lt:};
	\node [instr] (lt2) [below = of lt1] {cfi\_check 0x60} ;
	\node [instr] (lt3) [below = of lt2] {...} ;
	\node [instr] (lt4) [below = of lt3] {ret} ;

	\node [instr,fill=STblue!30] (gt1) [right = of trfoo8, xshift=1cm] {gt:};
	\node [instr] (gt2) [below = of gt1] {cfi\_check 0x70} ;
	\node [instr] (gt3) [below = of gt2] {...} ;
	\node [instr] (gt4) [below = of gt3] {ret} ;

	\node (fooentry)  [left = of foo2, xshift=-0.5cm] {};
	\node (fooexit)  [right = of foo9, xshift=0.5cm, yshift=0.5cm] {};
	\node (gtexit)  [right = of gt4, xshift=0.5cm] {};
	
	\draw [arrow, color=gray!50, ultra thick] (fooentry) -- (foo2);
	\draw [arrow, color=gray!50] (foo9) -|([shift={(5mm,0mm)}]foo9.east) -- (fooexit);
	\draw [arrow, color=STdarkgreen] (foo7) -|([shift={(5mm,0mm)}]foo7.east) -- ([shift={(-5.05mm,0mm)}]trfoo2.west)|-(trfoo2);
	\draw [arrow, color=STdarkgreen] (trfoo6) -|([shift={(5mm,0mm)}]trfoo6.east) -- ([shift={(-5.05mm,0mm)}]lt3.west)|-(lt3);
	\draw [arrow, color=STdarkgreen] (trfoo8) -|([shift={(5mm,0mm)}]trfoo8.east) -- ([shift={(-5.05mm,0mm)}]gt3.west)|-(gt3);
	
	\draw [arrow, color=STdarkgreen] (lt4) -|([shift={(5mm,-3.7cm)}]lt4.east) -- ([shift={(-5.05mm,-2.5cm)}]foo8.west)|-(foo8);
	\draw [color=STdarkgreen, ultra thick] (gt4) -- (gtexit);
	
\end{tikzpicture}

}
 	\caption{Application of \textit{trampoline} in \ac{EXCEC}}
 	\label{fig:excec_trampoline}
 \end{figure}

When an indirect call targets functions that are potentially called from multiple places, a common label-based approach would require to instrument all of these callers and callees with the same label.
This of course reduces the precision of the abstract \ac{CFG} used for \ac{CFI} enforcement.
However, \textit{trampolines} help to solve this problem by introducing a unique intermediate step between some indirect function call and its targets.
\autoref{fig:excec_trampoline} depicts the workings of an exemplary trampoline in \ac{EXCEC}.

When using this concept, the target of an indirect call is changed such that it points to its unique trampoline (\texttt{\_tr\_foo} in \autoref{fig:excec_trampoline}), while storing the actual target address of the call on the stack. 
This control flow transfer is protected with a unique label (\texttt{0x42} in the example). 
The trampoline itself contains a jump table with all valid call targets (\texttt{lt} and \texttt{gt} in this case) and checks whether the target address of the indirect function call, which is retrieved from the stack at the beginning of the trampoline, is contained in this table.
When some matching entry is found, the trampoline branches to the intended function but bypasses the \texttt{CFI\_CHECK} at its entry.
Such a check is not required here because the branch is only a direct jump now.
Otherwise, when no entry is found in the table, a \ac{CFI} violation is triggered by the \texttt{CFI\_CHECK} instruction at the end of the trampoline.
This instruction is only reached when all previous lookups in the table failed.
The jumps to a trampoline and from there further to the actual function are unidirectional, i.e., the function does not return to the trampoline but to its original call site.

Note that trampolines are only required for functions that can be called from multiple callers and are not generated otherwise.
For frequently visited indirect call sites with many targets the order of branches in the trampoline influences the overall performance.
Some kind of optimization, e.g., ordering them according to profiling information, can be applied to reduce runtime.

\subsubsection{Backward Edges}
We implement backward-edge \ac{CFI} enforcement with a shadow stack to provide optimal protection.
Unlike other concepts, we utilize existing RISC-V \texttt{JAL/JALR} and \texttt{RET} instructions for immediately manipulating the shadow stack.
This is possible in our setup because we can directly access the respective signals in the core's decoder.
By doing so we are able to implement protection of function returns with zero run-time and code size overhead.

Furthermore, we do not store full 32-bit addresses on the shadow stack but only the bits [18:1], which cuts memory demands of the stack almost in half.
This is reasonable because the remaining bits are not relevant in our environment.
First of all, instructions always start at even addresses in the RISC-V \ac{ISA}, meaning the \ac{LSB} is always 0.
In \mbox{PULPissimo}'s memory map the 13 \acp{MSB} have a fixed value because of the limited use of address space, which makes storing and comparing them meaningless.
This reduction to the relevant bits is of course only possible in our setup and might not be generally applicable.

Dedicated recursion handling is implemented for \ac{EXCEC} in order to reduce utilization of the shadow stack.
Configurable counters are attached to all stack entries to keep track of the recursion depth.
This allows for precise monitoring of both recursive and non-recursive function calls.

\ac{EXCEC} also supports \texttt{setjmp} calls similarly to \cite{christoulakis2016hcfi}.
To that end, we store the current shadow stack pointer when \texttt{setjmp} calls occur.
Later, when the stack is unwound because of a \texttt{longjmp} call, the \ac{CFI} module also unwinds the shadow stack to the index previously stored.

We have implemented custom \ac{CFI} management commands in addition to the instructions already presented in \autoref{tbl:cfi_instructions}.
\texttt{CFI\_ENABLE} and \texttt{CFI\_DISABLE} can be used to turn \ac{CFI} protection on, respectively off. 
This is used to exclude certain sections like the startup code.
Any \ac{CFI} instruction occurring while \ac{CFI} is disabled is handled as a \texttt{NOP} and has no further effect.
\texttt{CFI\_RESET} resets all internal \ac{CFI} registers and disables \ac{CFI}.
We use the reset, for example, when an \texttt{exit} occurs because the shadow stack would otherwise not be unwound.

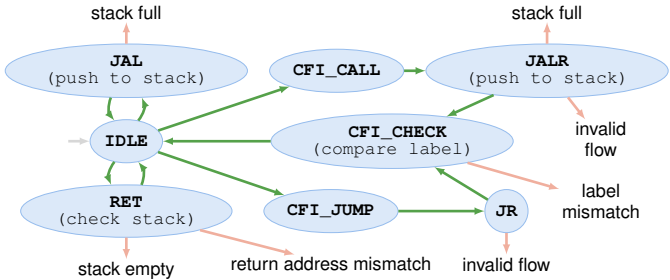
\begin{figure}[htb]
 	\centering
 	\resizebox{\linewidth}{!}{%
\begin{tikzpicture}[auto, node distance=1cm and 1cm, every loop/.style={},
	thick,main node/.style={circle,draw,font=\sffamily\Huge\bfseries},
	ball/.style={ellipse, inner sep=5pt, minimum width=2cm, minimum height=2cm, draw, font=\Huge\ttfamily}, > =latex, color=STblue!75, text=black, fill=STblue!50, align=center]
	\tikzset{edge/.style = {line width=3.5pt, ->}}
	
	\node [ball, fill=STblue!25] (IDLE) { \textbf{IDLE}};
	\node [ball, fill=STblue!25] (JAL) [above=of IDLE] { \textbf{JAL}\\[-0.5ex] (push to stack)};
	\node [ball, fill=STblue!25] (RET) [below=of IDLE] { \textbf{RET}\\[-0.5ex]  (check stack)};
	\node [ball, fill=STblue!25] (CFICALL) [right=of JAL, xshift=0cm] { \textbf{CFI\_CALL}};
	\node [ball, fill=STblue!25] (CFIJUMP) [right=of RET, xshift=0.45cm] { \textbf{CFI\_JUMP}};
	\node [ball, fill=STblue!25, node distance=5cm] (CFICHECK) [right=of IDLE] {\textbf{ CFI\_CHECK}\\[-0.5ex] (compare label)};
	\node [ball, fill=STblue!25] (JR) [right=of CFIJUMP, xshift=3cm] { \textbf{JR}};
	\node [ball, fill=STblue!25] (JALR) [right=of CFICALL] { \textbf{JALR}\\[-0.5ex]  (push to stack)};
	
	\node [draw=none, font=\sffamily\Huge] (JALSTACKFULL) [above=of JAL] {\Huge stack full};
	\node [draw=none, font=\sffamily\Huge] (RETSTACKEMPTY) [below=of RET] {\Huge stack empty};
	\node [draw=none, font=\sffamily\Huge] (JALRSTACKFULL) [above=of JALR] {\Huge stack full};
	\node [draw=none, font=\sffamily\Huge] (JRINVALIDFLOW) [below=of JR] {\Huge invalid flow};
	\node [draw=none, font=\sffamily\Huge] (JALRINVALIDFLOW) [below=of JALR, xshift=2.5cm] {\Huge invalid\\ \Huge flow};
	\node [draw=none, font=\sffamily\Huge] (CFICHECKMISMATCH) [below=of JALRINVALIDFLOW] {\Huge label\\ \Huge mismatch};
	\node [draw=none, font=\sffamily\Huge] (RETMISMATCH) [below=of CFIJUMP] {\Huge return address mismatch};
	\node [draw=none, font=\sffamily\Huge] (ENTRY) [left=of IDLE] {};
	
	\draw [edge, color=STdarkgreen!75] (IDLE) to [bend right] node[sloped, align=center, below]{} (JAL) ;
	\draw [edge, color=STdarkgreen!75] (JAL) to [bend right] node[sloped, align=center, below]{} (IDLE) ;
	
	\draw [edge, color=STdarkgreen!75] (IDLE) to [bend right] node[sloped, align=center, below]{} (RET) ;
	\draw [edge, color=STdarkgreen!75] (RET) to [bend right] node[sloped, align=center, below]{} (IDLE) ;
	
	\draw[edge, color=STdarkgreen!75] (IDLE) -- (CFICALL) node [midway] {};
	\draw[edge, color=STdarkgreen!75] (IDLE) -- (CFIJUMP) node [midway] {};
	\draw[edge, color=STdarkgreen!75] (CFICALL) -- (JALR) node [midway] {};
	\draw[edge, color=STdarkgreen!75] (CFIJUMP) -- (JR) node [midway] {};
	\draw[edge, color=STdarkgreen!75] (JALR) -- (CFICHECK) node [midway] {};
	\draw[edge, color=STdarkgreen!75] (JR) -- (CFICHECK) node [midway] {};
	\draw[edge, color=STdarkgreen!75] (CFICHECK) -- (IDLE) node [midway] {};
	
	\draw[edge, color=STred!75] (JAL) -- (JALSTACKFULL);
	\draw[edge, color=STred!75] (JALR) -- (JALRSTACKFULL);
	\draw[edge, color=STred!75] (RET) -- (RETSTACKEMPTY);
	\draw[edge, color=STred!75] (JR) -- (JRINVALIDFLOW);
	
	\draw[edge, color=STred!75] (JALR) -- (JALRINVALIDFLOW);
	\draw[edge, color=STred!75] (CFICHECK) -- (CFICHECKMISMATCH);
	\draw[edge, color=STred!75] (RET) -- (RETMISMATCH);

	\draw[edge, color=STgrey!75] (ENTRY) -- (IDLE) node [midway] {};
	
\end{tikzpicture}
}
 	\caption{Valid instruction sequences in \ac{EXCEC}}
 	\label{fig:excec_call_sequences}
 \end{figure}

The correct sequence of the instructions listed in \autoref{tbl:cfi_instructions} is enforced with an internal state machine.
We show all of such valid sequences in \autoref{fig:excec_call_sequences}.
In particular, the indirect control transfer instructions \texttt{JR} and \texttt{JALR} must be followed by \texttt{CFI\_CHECK}.
\texttt{JR} and \texttt{JALR} are always inherently preceded by \texttt{CFI\_JUMP} or \texttt{CFI\_CALL} due to the instrumentation.
\texttt{CFI\_CHECK} makes sure the labels match.
Similarly, \texttt{RET} compares the data from the top of the shadow stack with the return address.
Any deviations result in a \ac{CFI} violation, which triggers an exception in the core's pipeline controller that leads to immediate program termination if no exception handler recovers from this situation.
Instrumentation ensures that \texttt{CFI\_CALL} and \texttt{CFI\_JUMP} only occur before control flow transfers.
\texttt{CFI\_CHECK} with label \texttt{0x0} always triggers a \ac{CFI} exception.

Reacting upon such exceptions would make it possible to, for example, log the error cause and reset otherwise left dirty \ac{CFI} registers with the previously mentioned \texttt{CFI\_RESET} instruction.
In addition, the \ac{CFI} module tests whether the stack is already full before pushing a new entry
(or whether a bound recursion counter is exceeded), respectively whether it is empty before popping one.
While the former is technically no \ac{CFI} violation, the latter hints at an invalid control flow.

\subsubsection{Interrupts}
Unlike the referenced work on fine-grained \ac{CFI} enforcements we make sure that regular interrupt handling is not perturbed.
To that end, we enforce atomicity on all guarded control flow transitions by disabling interrupts temporarily while the \ac{CFI} state machine is active, i.e., between the \texttt{CFI\_CALL}/\texttt{CFI\_JUMP} and \texttt{CFI\_CHECK} instructions.
This guarantees consistency of \ac{EXCEC}'s internal state and enables the enforcement of \ac{CFI} even in \acp{ISR} at the cost of a few additional cycles of interrupt latency.
\section{Evaluation}
\label{sec:evaluation}

\subsection{Platform}

We implemented all of the \ac{CFI} schemes discussed in this work on the open-source PULP platform~\citeLinks[L][]{pulpPlatform}, 
which contains the fully synthesizable \mbox{PULPissimo} microcontroller architecture~\citeLinks[L][]{pulpissimo} 
as well as the associated RISC-V toolchain~\citeLinks[L][]{pulpRiscvToolchain} and SDK with platform-specific code~\citeLinks[L][]{pulpSdk}.
In our setup, the \mbox{PULPissimo} \ac{SoC} has one \SI{16}{MHz} \ac{CPU} core and a total of \SI{512}{KiB} of memory, which limits possible benchmark applications.
The system comes with two selectable 32-bit core variants, namely CV32E40P~\citeLinks[L][]{cv32e40p}  (formerly RI5CY~\cite{gautschiNearthresholdRISCVCoreDSP2017}) and Ibex~\citeLinks[L][]{ibex} (formerly Zero-riscy~\cite{schiavoneSlowSteadyWinsRace2017}).
We decided to use the CV32E40P core since it is the more capable default selection and bears a closer resemblance to commercial platforms that presumably are most likely to get additional security features (which is a monetary burden for \acp{SoC} based on very small cores).
CV32E40P is an in-order core with a 4-stage pipeline and implements the standardized RISC-V extensions for compressed instructions, integer multiplications/division and (optionally) floating point operations additionally to the base integer \ac{ISA}.
A PULP-specific extension is also supported, adding for example dedicated hardware loops and bit-wise operations.
The core does not come with caches but uses a prefetch buffer for instruction fetching.
For evaluation we used a Xilinx Zedboard~\citeLinks[L][]{zedboard}, which has the Zynq \mbox{Z-7020} \acs{SoC}-\acs{FPGA} installed.
We only used its \ac{PL} part that offers 53,200 \acp{LUT} and 106,400 \acp{FF}.
All benchmarks presented here result from executions on this \ac{FPGA} with the GCC options \texttt{-O2 -g0}.
We also used the \texttt{-fno-optimize-sibling-calls} flag in order to disable sibling- and tail recursive call optimizations, which are not compatible with \ac{CFI}.
This is because such calls violate the principle that function calls must return to their original call site.
Not doing so leads to \ac{CFI} inconsistencies on backward edges, e.g., with shadow stacks \cite{christoulakis2016hcfi}.
In this configuration, the \mbox{PULPissimo} \ac{SoC} with the CV32E40P core achieves a CoreMark~\citeLinks[L][]{coremark} score of 43.85 and 37.44 with and without the PULP extension (i.e., 2.74 and 2.34 CoreMarks/MHz), respectively, with GCC 7.1.1.

\subsection{Instrumentation}

All control flow transfers span a \ac{CFG} that is required for the instrumentation process.
\Ac{CFG} information is implicitly available for direct branches and function returns at build time.
Indirect jumps and calls, however, require explicit information.
The precision of the resulting \ac{CFG} directly translates to granularity of \ac{CFI} protection.
Extracting the possible set of targets of indirect control flow transfers in order to construct such a \ac{CFG} perfectly is (in general) undecidable~\cite{gonzalvezCaseIndirectJumpsSecure2019} and out of the focus of this work.

Therefore, we have manually prepared \ac{CFG} information on all indirect calls in the chosen benchmark applications.
This resembles what a \textit{perfect} compiler would do and thus might be more precise than automatically generated \acp{CFG}.
\Ac{CFG} information consists of a list of indirect control flow transfers identified by file name, function name and line number where they occur, along with a label for forward edge protection.
This approach is practicable because only five of our benchmark applications use indirect function calls at all in their actual code and none contains indirect jumps other than the ones generated with jump tables for switches.

To make full use of symbol information available at build time, we have implemented instrumentation by means of a GCC plugin.
GCC performs various phases called \textit{passes} and allows external plugins to register additional passes.
Our plugin is hooked in as one of the very last steps to be executed where we determine for each function if and where \ac{CFI} instructions need to be injected.

We use \ac{LTO} in order to enable unified optimization when linking all compilation units.
This allows us to also instrument the target-specific functions statically linked from the PULP SDK.
Unfortunately, the \texttt{libgcc} library that implements all compiler builtins (e.g., to handle arithmetic operations that the target processor does not support) cannot be properly built to support \ac{LTO} as of this time.
While this is not a fundamental flaw of our approach, any calls to functions of \texttt{libgcc} cannot be properly instrumented because they have to be linked unmodified without \ac{LTO}.
Therefore some functions, for example those declared by \texttt{math.h}, need to be excluded from \ac{CFI} instrumentation.
Otherwise only calls of respective functions would be instrumented but not their returns, which breaks backward edge protection concepts.
We add \texttt{NOPs} instead wherever possible to account for the missing instrumentations so that we achieve the same result in terms of run-time- and code size overhead in the end.
Our implementation of Intel CET particularly suffers from this limitation because its \ac{CFI} module automatically switches to a state where a specific instruction is expected upon every indirect jump and call.
We cannot inject these expected instructions into \texttt{libgcc} code, which forced us to disable \ac{CFI} enforcement for CET during our benchmarks.
This does, however, neither affect performance nor code size and thus does not harm comparability with other concepts because indirect function calls in \texttt{libgcc} are nowhere instrumented.

\subsection{Benchmarks}
Our evaluation is based on a set of 39 applications, most of which are part of respected benchmark suits.
To enable a meaningful evaluation and avoid any selection bias we include a wide range of diverse and commonly used programs for embedded systems tests.
The restriction on embedded applications is primarily necessary due to the memory constraints of our platform.
In most cases it was possible to port and execute whole suits.

A key component is CoreMark~\citeLinks[L][]{coremark}, which is an industry standard for benchmarking embedded CPUs.
We also use all programs of the Embench-IoT suite~\citeLinks[L][]{embenchIoT} and all single-threaded applications of UCB's RISC-V benchmarks~\citeLinks[L][]{riscvBenchmarks}.
Furthermore, we use a selection of small programs from the MiBench suite~\citeLinks[L][]{mibench}.
The excluded programs are not suitable for embedded platforms due to large problem sets or using file I/O.
In addition, we included three supplementary custom applications for dedicated recursion tests.

In total, five out of the 39 applications contain indirect function calls, which is particularly relevant for evaluating run-time overhead of forward edge \ac{CFI} protections.
While their focus on embedded applications might seem to limit the possible conclusions drawn, the distribution of instructions for control flow transfers is not vastly different in much bigger applications~\cite{AsmDB}.
However, with some work loads this can be significantly different:
For example, interpreters have a notoriously high amounts of indirect control transfers of about 5--10\% of all executed instructions~\cite{ertl2003interpreters}.

The selected benchmarks work with all of our hardware-based \ac{CFI} implementations, with two exceptions:
\texttt{slre} of the Embench-IoT suite contains nested recursive calls, which are not supported by HAFIX, and \texttt{towers} of the RISC-V benchmarks for HCFI, also because of problems with recursion.
We omit these two applications when computing aggregated results of the run-time overheads for all of our \ac{CFI} implementations.

The only changes to the original benchmarks' code are a unified approach for timing for all applications, 
which we implemented by reading the RISC-V \acp{CSR} for cycles and instructions before and after executing the benchmark's actual core algorithms, 
and some minor modifications such as replacements for library functions not provided by the PULP SDK.
In addition, we had to decrease the problem set size for \texttt{qsort} and \texttt{susan} (both from the MiBench suite) due to the memory size limitations of the platform.
For the same reason we have increased the stack size of applications from \SI{2}{KiB} to \SI{18}{KiB}.
Execution of multiple benchmarks would not be feasible otherwise.

\subsection{Implementation parameters}
\label{sec:parameters}

We first define the set of common parameters shown in \autoref{tbl:parameters} in order to enable a meaningful comparison of the hardware utilization overhead entailed by various \ac{CFI} concepts.
They specify the dimensions of all relevant memory elements and are employed in a consistent way throughout all implementations wherever applicable.
The values are based on our set of benchmarks but also consider constraints imposed by the limited hardware resources available on the \ac{FPGA}.

\begin{table}[htb]
	\centering
	\begin{tabular}{ l r }
		\toprule
			\textbf{Parameter} & \textbf{Value} \\
		
		\midrule
			\texttt{SHADOW\_STACK\_SIZE\ \ \ \ \ } & \ \ \ 128 \\
			\texttt{RECURSION\_DEPTH} & 128 \\
			\texttt{INDIRECT\_CALLS} & 64 \\
			\texttt{INDIRECT\_JUMPS} & 64 \\
			\texttt{INDIRECTLY\_CALLED} & 64 \\
			\texttt{NUM\_FUNCTIONS} & 1024 \\
			\texttt{SETJMP\_CALLS} & 8 \\

		\bottomrule
	\end{tabular}
	\caption{Common implementation parameters}
	\label{tbl:parameters}
\end{table}

The call depth of nested function calls never exceeds 50 in our applications.
This is particularly true for non-recursive programs, where call depth hardly exceeds 10.
The \texttt{SHADOW\_STACK\_SIZE} parameter used allows for some realistic headroom in other applications and bigger data sets.
\texttt{RECURSION\_DEPTH} defines the maximal supported recursion depth used in \ac{CFI} concepts with dedicated recursion handling using counters 
(with a width of 7 bits in the evaluation).

A value greater than the supported stack size is not meaningful, considering that some \ac{CFI} schemes do not specifically handle recursion at all.
\texttt{INDIRECT\_CALLS}, \texttt{INDIRECT\_JUMPS} and \texttt{INDIRECTLY\_CALLED} gives the numbers of indirect calls, jumps and indirectly called functions, respectively, and thus determine required label widths and matrix dimensions.
Those are based on observations regarding the number of indirect control flow transfers in our set of benchmark applications.
\texttt{NUM\_FUNCTIONS} specifies the total number of functions a program may contain.
This information is required, for example, to determine the dimensions of the Active Set.
Eventually we use \texttt{SETJMP\_CALLS} to define the number of distinct \texttt{setjmp} calls supported.
We note that none of the programs we use for evaluation contains such calls.
Where available, the respective hardware mechanisms have been enabled for completeness and verified by custom test cases.

The parameters presented here can be translated directly to \ac{FF} demand.
For instance, a common shadow stack requires 128x32 \acp{FF} on our 32-bit plaform when applying the \texttt{SHADOW\_STACK\_SIZE} we defined in in \autoref{tbl:parameters}.
Naturally, doubling the shadow stack size to support 256 nested function calls doubles its \ac{FF} requirements and also needs significant amounts of additional \acp{LUT} and \acp{MUX}.
This must be taken into account when discussing dimensions of every memory element.
\section{Results}
\label{sec:results}

Implementations of the previously presented \ac{CFI} approaches on a common platfrom made it possible to generate meaningful benchmark results.
In this section we first give an introductory overview of the protectional scope of the respective \ac{CFI} implementations and then present comparisons of the introduced overheads.

\subsection{Security}
The \ac{CFI} schemes presented in this work mostly support different \ac{CFI} protection mechanisms. 
We give a brief overview in \autoref{tbl:security}, although qualitative security evaluations are not our focus.
The table shows which control flow transfers are protected by each of the \ac{CFI} schemes.
\textit{Fine-grained \ac{CFG}} means that the respective approach is based on some precise \ac{CFG} and does not only protect control flow transfers in a way similar to Branch Regulation.
We refer again to \cite{de2017survey} and the respective papers for details on each of the \ac{CFI} approaches.

\begin{table}[h]
	\centering
  \begin{tabularx}{0.81\linewidth}{l*6{c}}
    \toprule
                         & \rot{FIXER}  & \rot{HAFIX} & \rot{HCFI} &  \rot{HECFI} & \rot{CET} & \rot{EXCEC} \\ 

    \midrule
      Function returns   & \fullcirc    & \fullcirc   &  \fullcirc & \fullcirc    & \fullcirc       & \fullcirc \\ 
      Indirect calls     & \fullcirc    & \emptycirc  &  \fullcirc & \fullcirc    & \fullcirc       & \fullcirc \\
      Indirect jumps     & \emptycirc   & \emptycirc  &  \emptycirc& \fullcirc    & \fullcirc       & \fullcirc \\ 
      Fine-grained CFG   & \fullcirc    & \emptycirc  &  \fullcirc & \fullcirc    & \emptycirc      & \fullcirc \\ 
      Protected Interrupts & \emptycirc & \emptycirc  &  \emptycirc & \emptycirc  & \fullcirc       & \fullcirc \\ 

    \bottomrule
  \end{tabularx}
  \caption{Overview of supported \ac{CFI} protection scope}
	\label{tbl:security}
\end{table}
 
The functional scope of \ac{CFI} schemes must be considered when interpreting the overheads on run-time performance, code size and hardware utilization, which we present in the following sections.
However, it is also important \textit{how} certain features are implemented.
For example, FIXER and HCFI offer the same functional scope in principle but use very different concepts for forward edge protection.
HAFIX on the other hand only protects function returns by means of an Active Set, thereby sacrificing some granularity.
\Ac{CET} only enforces \ac{CFI} based on overapproximated, i.e., coarse-grained, \acp{CFG} on forward edges, which is far less restrictive and thus prevents less attacks.
These important distinctions will also be reflected in the following evaluation results.

\subsection{Performance}

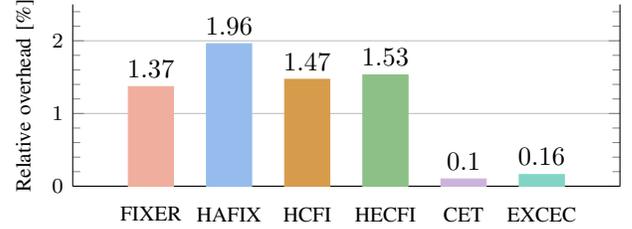
\begin{figure}[htb]
	\centering
	\begin{tikzpicture}
		\begin{axis}[ybar,
			width=\linewidth,
			height=4cm,
			bar width=.60cm,
			ymin=0,
			ymax=2.5,
			tick label style={font=\footnotesize},
			legend style={font=\footnotesize},
			label style={font=\footnotesize},      
			ylabel={Relative overhead [\%]},
			xmin=0,
        	xmax=7,
        	xtick={1,2,3,4,5,6},
        	x tick style={draw=none},
			xticklabels from table={\datatabletransposed}{name},
			every axis plot/.append style=fill,
			area legend,
			point meta=rawy,
			nodes near coords,
			nodes near coords align=vertical,
			nodes near coords style={black, /pgf/number format/fixed},
			ymajorgrids,
    		minor y tick num=4,
			axis x line*=bottom,
          	bar shift=0pt,
			]

			\addplot[STred!75,fill=STred!75] coordinates {(1,1.37)} ;
			\addplot[STblue!75,fill=STblue!75] coordinates {(2,1.96)} ;
			\addplot[STdarkorange!75,fill=STdarkorange!75] coordinates {(3,1.47)} ;
			\addplot[STdarkgreen!50,fill=STdarkgreen!50] coordinates {(4,1.53)} ;
			\addplot[STviolet!75,fill=STviolet!75] coordinates {(5,0.10)} ;
			\addplot[STturquoise!75,fill=STturquoise!75] coordinates {(6,0.16)} ;
			
		\end{axis}
	\end{tikzpicture}
	\caption{Relative runtime performance overhead}
	\label{fig:performance_overview}
\end{figure}

Arguably the most important aspect when comparing \ac{CFI} schemes is their impact on the runtime of applications.
\autoref{fig:performance_overview} shows the relative performance overhead for all \ac{CFI} implementations as an average over all benchmark applications.
\Ac{CET} and \ac{EXCEC} each introduce no more than 0.16\% overhead, which is due to the fact that these two concepts do not require additional instructions for backward edge protection but extend semantics of existing branch instructions instead.
Quite the opposite can be seen for HAFIX with an extra of 1.96\% because every function call involves at least three additional instructions, which is more than any other concept. 
FIXER, HCFI and HECFI follow a similar approach for backward edge protection, what explains why their results are close.
Forward edge protection, which is required less frequently, accounts for minor differences only.
However, applications with a large share of indirect function calls represent a worst-case scenario for most \ac{CFI} schemes:
Their different instrumentation concepts lead to an overhead of up to 4 \ac{CFI} instructions for every indirect call.
Schemes, that offer a very fine-grained protection by using trampolines can even add a multiple of this number in case they are needed.

\pgfplotstableread[col sep = semicolon]{diagrams/performance_overview.csv}\datatable
\pgfplotstabletranspose[colnames from=name, input colnames to=name]\datatabletransposed{\datatable}

For some combinations of \ac{CFI} schemes and benchmark applications the \ac{CFI} variants actually resulted in a lower execution time compared to a run without \ac{CFI} enforcement.
Instrumentation with \ac{CFI} instructions naturally changes the alignment of symbols within the \texttt{.text} section, which in individual cases even leads to improved performance.
A significant share of cycles, which make up these effects, can be traced to stalls resulting from non-ideal instruction fetches from memory.
For that reason we consider the number of executed instructions instead of the cycles required for execution throughout the evaluation.
Instruction counts are not affected by the above mentioned phenomenon.
Also, all run-time overheads presented here represent the average overhead for executions \textit{with} and \textit{without} using the PULP extension available on our platform.
As we later discuss, these account for considerable differences in individual cases, which are not strictly related to \ac{CFI} enforcement.

The detailed effects of \ac{CFI} protection on runtime for all \ac{CFI} schemes are shown in \autoref{tbl:performance_list}.
The numbers represent the relative overhead compared to a baseline without \ac{CFI} enforcement.
Indirect function calls make up only a small minority of forward control transfers and are contained in only 5 of our benchmark applications:
In \texttt{coremark} indirect function calls make up 15.4\% of all function calls, in \texttt{picojpeg} 0.1\%, in \texttt{wikisort} 25.7\%, and in \texttt{bitcount} 61.2\%.
The \texttt{sglib} benchmark contains indirect call instructions but they are never executed due to dynamic checks.
Indirect jumps are only marginally used for jump tables in \texttt{picojpeg} and \texttt{qrduino}.

\begin{table}[htb]
	\centering
  \begin{tabularx}{\linewidth}{llcccccc}
    \toprule
        \multicolumn{2}{c}{Benchmark} & \rot{FIXER}  & \rot{HAFIX} & \rot{HCFI} &  \rot{HECFI} & \rot{CET}       & \rot{EXCEC} \\ 

    \midrule

        & {coremark\textsuperscript{\dag}} & 1.42 & 2.13 & 1.53 & 1.64 & 0.11 & 0.22 \\
        \midrule

        \parbox[t]{0mm}{\multirow{19}{*}{\rotatebox[origin=c]{90}{Embench IoT}}} 
        & {aha-mont64} & 0.06 & 0.08 & 0.06 & 0.06 & 0.00 & 0.00 \\
        & {crc32} & 0.00 & 0.00 & 0.00 & 0.00 & 0.00 & 0.00 \\
        & {cubic} & 0.31 & 0.31 & 0.31 & 0.31 & 0.00 & 0.00 \\
        & {edn} & 0.00 & 0.00 & 0.00 & 0.00 & 0.00 & 0.00 \\
        & {huffbench} & 0.08 & 0.13 & 0.08 & 0.08 & 0.00 & 0.00 \\
        & {matmult-int} & 0.00 & 0.00 & 0.00 & 0.00 & 0.00 & 0.00 \\
        & {minver} & 0.65 & 0.98 & 0.65 & 0.65 & 0.00 & 0.00 \\
        & {nbody} & 1.07 & 1.07 & 1.07 & 1.07 & 0.00 & 0.00 \\
        & {nettle-aes} & 0.01 & 0.01 & 0.01 & 0.01 & 0.00 & 0.00 \\
        & {nettle-sha256} & 0.07 & 0.10 & 0.07 & 0.07 & 0.00 & 0.00 \\
        & {nsichneu} & 0.00 & 0.00 & 0.00 & 0.00 & 0.00 & 0.00 \\
        & {picojpeg\textsuperscript{\dag\ /\ \dag\dag}} & 1.14 & 1.71 & 1.14 & 1.17 & 0.03 & 0.06 \\
        & {qrduino\textsuperscript{\dag\ /\ \dag\dag}} & 0.17 & 0.26 & 0.17 & 0.18 & 0.00 & 0.00 \\
        & {sglib\textsuperscript{\dag}} & 2.99 & 4.48 & 3.11 & 3.11 & 0.12 & 0.12 \\
        & {slre\textsuperscript{*}} & 3.79 & n/a & 3.79 & 3.79 & 0.00 & 0.00 \\
        & {st} & 1.56 & 1.56 & 1.56 & 1.56 & 0.00 & 0.00 \\
        & {statemate} & 0.00 & 0.00 & 0.00 & 0.00 & 0.00 & 0.00 \\
        & {ud} & 1.84 & 1.89 & 1.84 & 1.84 & 0.00 & 0.00 \\
        & {wikisort\textsuperscript{\dag}} & 1.98 & 2.44 & 2.26 & 2.54 & 0.28 & 0.55 \\
        \midrule

        \parbox[t]{0mm}{\multirow{9}{*}{\rotatebox[origin=c]{90}{UCB RISC-V}}} 
        & {dhrystone} & 5.17 & 8.12 & 5.17 & 5.17 & 0.00 & 0.00 \\
        & {median} & 0.00 & 0.00 & 0.00 & 0.00 & 0.00 & 0.00 \\
        & {multiply} & 0.00 & 0.00 & 0.00 & 0.00 & 0.00 & 0.00 \\
        & {qsort} & 0.00 & 0.00 & 0.00 & 0.00 & 0.00 & 0.00 \\
        & {rsort} & 0.00 & 0.00 & 0.00 & 0.00 & 0.00 & 0.00 \\
        & {sort} & 0.00 & 0.00 & 0.00 & 0.00 & 0.00 & 0.00 \\
        & {spmv} & 1.43 & 1.43 & 1.43 & 1.43 & 0.00 & 0.00 \\
        & {towers\textsuperscript{*}} & 5.52 & 8.28 & n/a & 5.52 & 0.00 & 0.00 \\
        & {vvadd} & 0.00 & 0.00 & 0.00 & 0.00 & 0.00 & 0.00 \\
        \midrule
        
        \parbox[t]{0mm}{\multirow{7}{*}{\rotatebox[origin=c]{90}{MiBench}}} 
        & {basicmath} & 0.28 & 0.28 & 0.28 & 0.28 & 0.00 & 0.00 \\
        & {bitcount\textsuperscript{\dag}} & 5.95 & 8.93 & 8.93 & 10.75 & 2.98 & 4.80 \\
        & {dijkstra} & 0.15 & 0.23 & 0.15 & 0.15 & 0.00 & 0.00 \\
        & {fft} & 0.66 & 0.66 & 0.66 & 0.66 & 0.00 & 0.00 \\
        & {qsort\textsuperscript{\ddag}} & 0.76 & 1.15 & 0.76 & 0.76 & 0.00 & 0.00 \\
        & {stringsearch} & 0.00 & 0.00 & 0.00 & 0.00 & 0.00 & 0.00 \\
        & {susan\textsuperscript{\ddag}} & 0.08 & 0.08 & 0.08 & 0.10 & 0.00 & 0.00 \\
        \midrule
        
        \parbox[t]{0mm}{\multirow{3}{*}{\rotatebox[origin=c]{90}{custom}}} 
        & {factorial} & 14.88 & 22.31 & 14.88 & 14.88 & 0.00 & 0.00 \\
        & {nqueens} & 0.77 & 1.15 & 0.77 & 0.77 & 0.00 & 0.00 \\
        & {tak} & 7.26 & 10.89 & 7.26 & 7.26 & 0.00 & 0.00 \\

    \midrule
    \midrule

        & Maximum & 14.88 & 22.31 & 14.88 & 14.88 & 2.98 & 4.80\\
        & Average & 1.37 & 1.96 & 1.47 & 1.53 & 0.20 & 0.16 \\
        & Median & 0.17 & 0.26 & 0.17 & 0.18 & 0.00 & 0.00\\
        
    \bottomrule\\[-1.5ex]
        \multicolumn{8}{r}{\textsuperscript{*} excluded from aggregated results due to missing support}\\
        \multicolumn{8}{r}{contains indirect function calls\textsuperscript{\dag} or jumps\textsuperscript{\dag\dag}}\\
        \multicolumn{8}{r}{\textsuperscript{\ddag} reduced problem set}\\

	\end{tabularx}
	\caption{Detailed runtime performance overhead [\%].}
	\label{tbl:performance_list}
\end{table}

The averaged numbers shown before in \autoref{fig:performance_overview} hide significant deviations of particular benchmarks.
Therefore, we specifically analyze the remarkable results of some benchmark applications as seen in \autoref{fig:performance_details} to make them more comprehensible.
Applications with hardly any control flow transfers of course entail fewer \ac{CFI} instructions.
This is, for example, the case for the \texttt{basicmath} benchmark, where the timed portion of the program only contains a small number of function calls.
The opposite is true for recursive applications like our \texttt{factorial} implementation, where calls make up about 7.5\% of all instructions.
For instance, HCFI and FIXER both require two additional instructions for protecting each direct function call, or rather its return, resulting in their 15\% run-time overhead for \texttt{factorial}.

Intel CET and \ac{EXCEC} do not require instrumentation of direct function calls, meaning they introduce no direct run-time overhead at all for most applications.
However, their performance numbers for \texttt{coremark} bear witness to the fact that it contains indirect branches.
For all variants of \ac{CFI} enforcements the results for \texttt{coremark} are relatively close to the average values shown in \autoref{fig:performance_details}.
Indirect function calls also occur in \texttt{bitcount}, only now these make up for almost two thirds of all calls.
The \texttt{dhrystone} benchmark contains particularly many direct function calls, as can be seen from the run-time overhead introduced by FIXER, HAFIX, HCFI and HECFI.

\pgfplotstableread[col sep = semicolon]{diagrams/performance_details.csv}\datatable

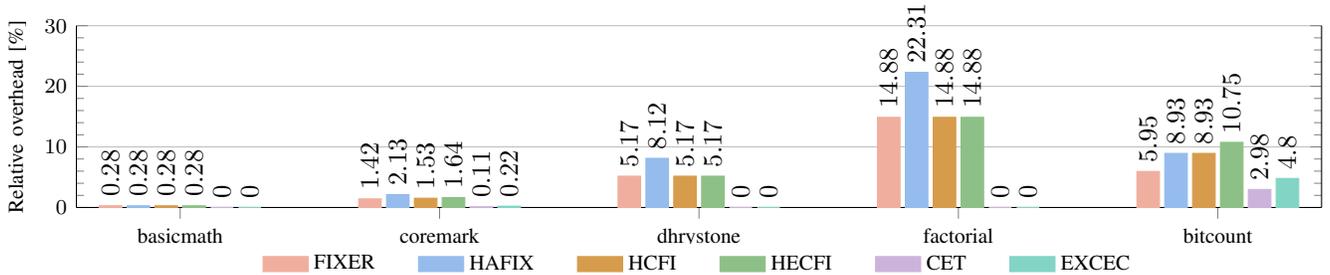
\begin{figure*}[htb]
	\centering
	\begin{tikzpicture}
		\begin{axis}[ybar,
			width=\linewidth,
			bar width=.3cm,
			legend style={
				legend columns=-1,
				at={(axis description cs:0.5,-0.2)},
				anchor=north,
				draw=none,
				/tikz/every even column/.append style={column sep=0.5cm}
			},
			ymin=0,
			ymax=30,
			tick label style={font=\footnotesize},
			legend style={font=\footnotesize},
			label style={font=\footnotesize},      
			ylabel={Relative overhead [\%]},
			height=4cm,   
			xtick=data,
			xticklabels from table={\datatable}{name},
			cycle list name=color list,
			every axis plot/.append style=fill,
			area legend,
			point meta=rawy,
			nodes near coords,
			nodes near coords align=vertical,
			nodes near coords style={/pgf/number format/fixed},
			every node near coord/.append style={black, rotate=90, anchor=west},
			ymajorgrids,
    		minor y tick num=4,
			axis x line*=bottom,
	  ]

	\addplot[STred!75,fill=STred!75] table [x expr=\coordindex, y=FIXER]\datatable;
	\addplot[STblue!75,fill=STblue!75] table [x expr=\coordindex, y=HAFIX]\datatable;
	\addplot[STdarkorange!75,fill=STdarkorange!75] table [x expr=\coordindex, y=HCFI]\datatable;
	\addplot[STdarkgreen!50,fill=STdarkgreen!50] table [x expr=\coordindex, y=HECFI]\datatable;
	\addplot[STviolet!75,fill=STviolet!75] table [x expr=\coordindex, y=CET]\datatable;
	\addplot[STturquoise!75,fill=STturquoise!75] table [x expr=\coordindex, y=EXCEC]\datatable;

	\legend{FIXER, HAFIX, HCFI, HECFI, CET, EXCEC}
	\end{axis}
	\end{tikzpicture}

\caption{Comparison of runtime performance overhead for conspicuous benchmark applications}
\label{fig:performance_details}
\end{figure*}

The run-time performance overhead of HAFIX and HECFI measured by us is more or less identical to what is stated in the original papers of the two approaches.
Minor differences for the HCFI numbers presumably result from the set of benchmark applications.
However, the performance indicators for FIXER need to be put into perspective because the original concept as described in \cite{de2019fixer} uses more instructions for instrumenting control flow transfers than our implementation does.
We described the reasons before when introducing FIXER.
The benchmarks used for evaluation in the original paper mostly only contain few function calls, so \ac{CFI} protection is naturally very cheap there.
The more extensive set of benchmarks used in this work would result in a significantly higher run-time overhead if instrumenting in the way proposed in the original concept.

The \mbox{PULPissimo} platform we used for all evaluations extends the RISC-V \ac{ISA} with an optional PULP extension for platform specific instructions like hardware loops.
These have a very positive effect on performance in many cases. 
For example, 92\% more instructions are executed for the \texttt{edn} benchmark when compiling the application \textit{without} the PULP extension and without any \ac{CFI} enforcement.
Minor impacts thereof are also reflected in the run-time performance overhead introduced by our \ac{CFI} schemes though.
We take this into account by always averaging run-time overheads of executions with and without PULP extensions.

In conclusion it can be said that all \ac{CFI} schemes are very performant on average and differences are rather small.
Noticeable differences can be seen in certain individual examples though, especially for applications containing an above-average number of (indirect) function calls.
In some extreme cases it might be beneficial to benchmark the set of actual applications to be executed to determine the most suitable \ac{CFI} enforcement scheme.

\subsection{Hardware Utilization}

Our \ac{CFI} implementations affect the maximum frequency of the overall hardware design to a small extent.
The vanilla \mbox{PULPissimo} platform with the CV32E40P core can be synthesized for the ZedBoard with \SI{16}{MHz} without any timing violations and still some margin available.
Most implementations have negligible impact on the timing but FIXER decreases the available margin noticeably.
In our implementation, of all concepts FIXER also introduces the most levels of logic (i.e., the highest amount of combinational elements between any two synchronous points) in its \ac{CFI} module.
The reason is that its significant amount of memory also requires a lot of associated logic, especially for the policy matrix lookup.

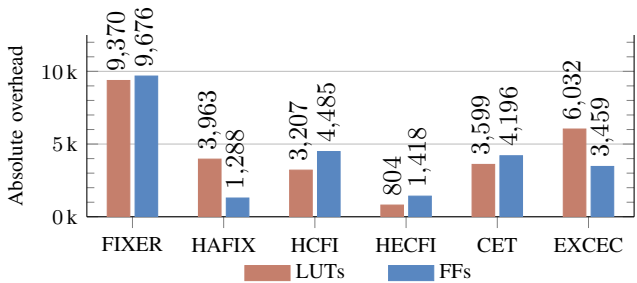
\begin{figure}[htb]
	\centering
	\begin{tikzpicture}
		\begin{axis}[ybar,
			width=\linewidth,
			bar width=.3cm,
			legend style={
				legend columns=-1,
				at={(axis description cs:0.5,-0.2)},
				anchor=north,
				draw=none,
				/tikz/every even column/.append style={column sep=0.5cm}
			},
			yticklabel style={
				/pgf/number format/fixed,
				/pgf/number format/precision=0
			},
			yticklabel = {
				\pgfmathparse{\tick/1000}
				\pgfmathprintnumber{\pgfmathresult}\,k
			},
			scaled y ticks=false,
			ymin=0,
			ymax=12500,
			tick label style={font=\footnotesize},
			legend style={font=\footnotesize},
			label style={font=\footnotesize},      
			ylabel={Absolute overhead},
			height=4cm,   
			xtick=data,
			xticklabels from table={\datatabletransposed}{name},
			cycle list name=color list,
			every axis plot/.append style=fill,
			area legend,
			point meta=rawy,
			nodes near coords,
			nodes near coords align=vertical,
			nodes near coords style={/pgf/number format/fixed},
			every node near coord/.append style={black, rotate=90, anchor=west},
			ymajorgrids,
    		minor y tick num=4,
			axis x line*=bottom,
			]
			
			\addplot[STdarkred!65,fill=STdarkred!65] table [x expr=\coordindex, y=lut]\datatabletransposed;
			\addplot[STdarkblue!65,fill=STdarkblue!65] table [x expr=\coordindex, y=ff]\datatabletransposed;
			
			\legend{LUTs, FFs}
		\end{axis}
	\end{tikzpicture}
	\caption{Additionally required hardware resources}
	\label{fig:utilization_overview}
\end{figure}

When evaluating hardware utilization overhead, we differentiate between \acp{LUT} and \acp{FF}.
While the former is more sensitive to implementation details, the latter mostly depends on the dimensions used for the various memory elements.
We defined the parameters used for such components in \autoref{sec:parameters}.
\autoref{fig:utilization_overview} shows the resulting utilization overhead for all of our \ac{CFI} variants.
Note that absolute numbers are shown here instead of relative increases to ease future comparisons with other platforms.
We discuss how the respective \acp{FF} demands are composed below.
Unfortunately, additional \acp{LUT} can not be trivially assigned.

\pgfplotstableread[col sep = semicolon]{diagrams/utilization_overview.csv}\datatable
\pgfplotstabletranspose[colnames from=name, input colnames to=name]\datatabletransposed{\datatable}

HAFIX does not offer forward edge protection and uses an Active Set for \ac{CFI} enforcement on function returns.
This of course results in relatively low hardware demands. 
FIXER on the other hand comes with both a shadow stack and a policy matrix, which are both rather expensive design elements.
The required address decoder, that translates 32-bit addresses to matrix indices, is also to be mentioned here.
The remaining approaches all use some sort of labels for forward edge protection, which only requires a negligible number of additional \acp{FF}.
\Ac{CFI} enforcement on backward edges is implemented with variations of shadow stacks in all of these concepts, only varying the stack width.
HCFI and \ac{CET} store full 32-bit addresses, \ac{EXCEC} 18 relevant address bits and HECFI even only stores short labels on its shadow stack, the width of which depends on the \texttt{NUM\_FUNCTIONS} parameter from \autoref{tbl:parameters}.
It is therefore obvious to consider some sort of reduced stack width because the full 32-bit address space is seldom used for instructions (especially in embedded systems).
Without this optimization \ac{EXCEC} would require 10,070 \acp{LUT} and 5,318 \acp{FF}, i.e., about 67\% and 54\% more overhead.

Most parts of \ac{FF} demand can be directly deduced from the dimensions of required memory modules.
For example, the 64x64 policy matrix used in FIXER requires 4096 \acp{FF} with the parameters defined in \autoref{tbl:parameters}.
The associated address decoder, which we implemented as a 64x18 lookup table, accounts for 1152 \acp{FF} and the 128x32 bit shadow stack is built from another 4096 \acp{FF}.
Other small registers make up the difference to the sum shown in \autoref{fig:utilization_overview}.
EXCEC's reduced shadow stack width results in a lower requirement of 2304 \acp{FF} (128x18).

It becomes apparent that approaches, which are based on some sort of policy matrix, are hard to implement and very expensive.
Matrix dimensions are defined by the numbers of supported callers and callees.
Doubling these parameters would quadruple the number of required \acp{FF}.
In contrast, label-based concepts hardly require any resources but still offer a good \ac{CFG} precision when being used with fine-grained labels or in combination with trampolines.

A direct comparison with the hardware utilization overhead stated in the original papers of the respective \ac{CFI} concepts is hardly meaningful:
The dimensions of various memory elements are mostly unknown and the different platforms used for evaluations vary greatly.
For example, about 2.5\% of additionally required \acp{FF} are reported for both HAFIX and HCFI and a general area overhead of 2.9\% is specified for FIXER.
These numbers differ significantly from ours -- sometimes even by a magnitude.

\subsection{Code Size}

The effects of \ac{CFI} instrumentation on the \texttt{.text} section size are shown in \autoref{fig:size_overview}.
Note that the figure shows a logarithmic scale in order to better represent the differences.  
In general, great outliers can be seen for all \ac{CFI} implementations.
Particularly small code size overheads are introduced by applications with only few function calls, and vice versa.

FIXER and HCFI show an almost identical overhead because the two concepts follow a very similar approach when it comes to instrumenting direct branches.
Both add one instruction each to the caller and callee for every function call.
The Active Set approach used in HAFIX on the other hand requires instrumentation of every function, regardless how often it is invoked, and an additional instruction for every call.
HECFI follows a quite expensive approach for backward edge protection in terms of code size by injecting the instruction used for popping from the shadow stack into the caller, i.e.,\ at the position where the function call returns to.
Other approaches like FIXER and HCFI instead instrument the callee's return, so only one instruction is added no matter how often the respective function is called.
The instrumentation concept used for HECFI therefore results in the same number of executed instructions, but the code size is significantly larger.
\Ac{CET} and \ac{EXCEC} stand out positively because neither of them needs to instrument direct function calls.
Only minor differences in the results stem from the instrumentation of indirect control flow transfers, which only account for a very small share of all control flow transfers.

\pgfplotstableread[col sep = semicolon]{diagrams/size_overview.csv}\datatable
\pgfplotstabletranspose[colnames from=name, input colnames to=name]\datatabletransposed{\datatable}

\tikzset{
	col/.style={
		mark=*,
		draw=#1,
		fill=#1,
		every mark/.append style={fill=#1}
	}
}

\pgfplotscreateplotcyclelist{graph_colors}{%
	col={STred!75}\\
	col={STgreen!75}\\
	col={STdarkorange!75}\\
	col={STblue!75}\\
	col={STviolet!75}\\
	col={STturquoise!75}\\
}

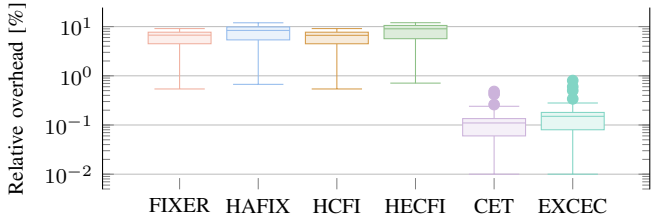
\begin{figure}[htb]
	\centering
	\begin{tikzpicture}
		\begin{axis}[
				boxplot/draw direction = y,
				width=\linewidth,
				height=4cm,
				bar width=.50cm,
				x axis line style = {opacity=0},
				axis x line* = bottom,
				enlarge y limits,
				ymajorgrids,
				tick label style={font=\footnotesize},
				label style={font=\footnotesize},      
				xtick={1,2,3,4,5,6}, 
				xticklabels from table={\datatabletransposed}{name},
				cycle list name=graph_colors,
				ymajorgrids,
				minor y tick num=4,
				axis x line*=bottom,
    			ymode=log,
				ylabel={Relative overhead [\%]},
				every box/.style={ultra thick},
				every whisker/.style={ultra thick},
				every median/.style={ultra thick},
			]

			\addplot+[boxplot, STred!75,fill=STred!15] table [x expr=\coordindex, y=FIXER]\datatable;
			\addplot+[boxplot, STblue!75,fill=STblue!15] table [x expr=\coordindex, y=HAFIX]\datatable;
			\addplot+[boxplot, STdarkorange!75,fill=STdarkorange!15] table [x expr=\coordindex, y=HCFI]\datatable;
			\addplot+[boxplot, STdarkgreen!50,fill=STdarkgreen!15] table [x expr=\coordindex, y=HECFI]\datatable;
			\addplot+[boxplot, STviolet!75,fill=STviolet!15] table [x expr=\coordindex, y=CET]\datatable;
			\addplot+[boxplot, STturquoise!75,fill=STturquoise!15] table [x expr=\coordindex, y=EXCEC]\datatable;

		\end{axis}
	\end{tikzpicture}
	\caption{Code size overhead (log scale)}
	\label{fig:size_overview}
\end{figure} %

\subsection{Summary}
\begin{figure}[htb]
	\centering
	\begin{tikzpicture}
		\begin{axis}[ybar,
			width=\linewidth,
			bar width=.25cm,
			legend style={
				legend columns=-1,
				at={(axis description cs:0.5,-0.2)},
				anchor=north,
				draw=none,
				/tikz/every even column/.append style={column sep=0.5cm}
			},
			yticklabel style={
				/pgf/number format/fixed,
				/pgf/number format/precision=0
			},
			scaled y ticks=false,
			ymin=0,
			ymax=40,
			tick label style={font=\footnotesize},
			legend style={font=\footnotesize},
			label style={font=\footnotesize},      
			ylabel={Relative overhead [\%]},
			height=4cm,   
			xtick=data,
			xticklabels from table={\datatabletransposed}{name},
			cycle list name=color list,
			every axis plot/.append style=fill,
			area legend,
			point meta=rawy,
			ymajorgrids,
    		minor y tick num=4,
			axis x line*=bottom,
			enlarge x limits={abs=0.7},
			]
			
			\addplot[STviolet!65,fill=STviolet!65] table [x expr=\coordindex, y=runtime]\datatabletransposed;
			\addplot[STdarkblue!65,fill=STdarkblue!65] table [x expr=\coordindex, y=size]\datatabletransposed;
			\addplot[STdarkgreen!65,fill=STdarkgreen!65] table [x expr=\coordindex, y=utilization]\datatabletransposed;
			
			\legend{Runtime, Code Size, \ac{HW}~Utilization}
		\end{axis}
	\end{tikzpicture}
  	\caption{Overview of relative overheads}
	\label{fig:summary_overview}
\end{figure}
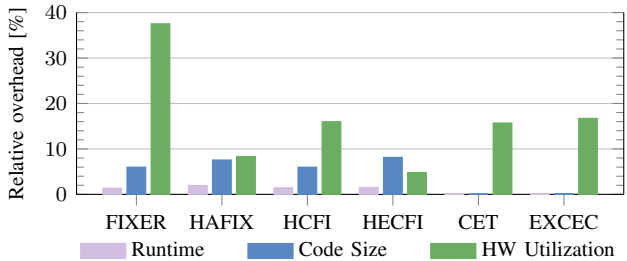

We present a final summary of the impact of \ac{CFI} enforcement on the previously described aspects performance, binary size and hardware utilization in \autoref{fig:summary_overview}.
For clarity the figure shows the relative overhead of hardware utilization as average of \acp{LUT} and \acp{FF} increase.
When ignoring security aspects then Intel CET and \ac{EXCEC} appear equivalent in this very simplified representation.
However, \ac{EXCEC} enforces a much more fine-grained \ac{CFI}.
The biggest outlier over all dimensions is FIXER's hardware utilization, which is caused by its expensive policy matrix.
Apart from that, the proportionally small performance overheads of all implementation seem to be almost negligible in comparison.

\pgfplotstableread[col sep = semicolon]{diagrams/summary_overview.csv}\datatable
\pgfplotstabletranspose[colnames from=name, input colnames to=name]\datatabletransposed{\datatable}

\section{Conclusion}

In this work we present an extensive evaluation of various state-of-the-art \ac{CFG}-based hardware-assisted \ac{CFI} schemes.
In addition, we unify the most promising concepts and ideas into a novel and precise \ac{CFI} approach called \acf{EXCEC}.

The elaborate set of benchmarks that has been ported to as well as the usage of a single platform allows for a fair, transparent and thorough evaluation.
The instrumentation is carried out largely automatically by a compiler plugin without the need to change the latter directly.
The results reveal the biggest differences in hardware costs and many implementations require a non-negligible amount of additional instructions inserted to the binaries increasing their size.

Our own protection scheme is shown to be more efficient in terms of run-time performance and binary size while involving an average hardware utilization overhead and enforcing a fine-grained \ac{CFG}.
We prove with this approach that fine-grained \ac{CFI} enforcement can be achieved with even less overheads than previously demonstrated, even if peculiar details (e.g., \texttt{setjmp}) are handled.
It follows that here is no need to resort to less precise and most often even costlier solutions as long as the \ac{CFG} is available.

We have released all our implementation artifacts including the Verilog hardware descriptions, GNU Binutils, and the GCC plugin as open-source in Git repositories~\cite{excec_repo}.

\subsection{Future Work}

The focus of this paper lies on benchmark applications written in C because of constraints imposed by the PULP SDK.
However, an extension of our GCC plugin for C++ support would allow us to investigate the challenges to \ac{CFI} enforcement posed by object-oriented languages.

Our instrumentation is based on partially manually constructed \acp{CFG}, which is practicable for academic evaluation but lacks broad applicability.
A combination of our instrumentation with some framework for extracting \ac{CFG} information from source code would alleviate this problem.

The modularity of our hardware mechanisms opens up opportunities to examine the effects of \ac{CFI} enforcement on other platforms, e.g., out-of-order pipelines, multi-core \acp{CPU} or alternative \acp{ISA}.
While the protection of the \ac{CFI} state (e.g., shadow stack contents and \ac{FSM} state) from deliberate direct manipulation in the current scheme is an important security feature in embedded systems without layered privileges it also prohibits an effective implementation of preemptive multitasking.
Providing privileged instructions to access this internal data would allow \pgls{OS} to spill the state to memory during context switches.
Similarly, this could raze the upper limit on call depths and recursions.
The way \ac{EXCEC} handles interrupts could be applied for other asynchronous events such as (Unix) signals.

All of the above combined would also allow for implementations and evaluations on larger systems including \pgls{OS} since there are no fundamental obstacles to do so.
Limiting \ac{CFI} enforcements to selected processes only would enable a gradual migration for legacy applications.
In such systems we predict an increased overhead due to additional indirect jumps, e.g., for dynamic linking, and from the bigger process metadata but this needs to be verified.
 \section*{Acknowledgements}
This work has been supported by the Doctoral College Resilient Embedded Systems, which is run jointly by the TU Wien's Faculty of Informatics and the UAS Technikum Wien.
The authors would also like to thank Andreas Steininger and all other reviewers of the manuscript for their helpful feedback.
\bibliographystyle{IEEEtranS}
\bibliography{}

\begin{thebibliography}{10}
\providecommand{\href}[2]{#1}
\providecommand{\url}[1]{#1}
\csname url@samestyle\endcsname
\providecommand{\newblock}{\relax}
\providecommand{\bibinfo}[2]{#2}
\providecommand{\BIBentrySTDinterwordspacing}{\spaceskip=0pt\relax}
\providecommand{\BIBentryALTinterwordstretchfactor}{4}
\providecommand{\BIBentryALTinterwordspacing}{\spaceskip=\fontdimen2\font plus
\BIBentryALTinterwordstretchfactor\fontdimen3\font minus
  \fontdimen4\font\relax}
\providecommand{\BIBforeignlanguage}[2]{{%
\expandafter\ifx\csname l@#1\endcsname\relax
\typeout{** WARNING: IEEEtran.bst: No hyphenation pattern has been}%
\typeout{** loaded for the language `#1'. Using the pattern for}%
\typeout{** the default language instead.}%
\else
\language=\csname l@#1\endcsname
\fi
#2}}
\providecommand{\BIBdecl}{\relax}
\BIBdecl

\bibitem{pulpPlatform}
\BIBentryALTinterwordspacing
``{PULP Platform}.''  \url{https://www.pulp-platform.org/}
\BIBentrySTDinterwordspacing

\bibitem{pulpissimo}
\BIBentryALTinterwordspacing
``{PULPissimo}.''  \url{https://github.com/pulp-platform/pulpissimo}
\BIBentrySTDinterwordspacing

\bibitem{pulpRiscvToolchain}
\BIBentryALTinterwordspacing
``{PULP RISC-V GNU Toolchain}.''
  \url{https://github.com/pulp-platform/pulp-riscv-gnu-toolchain}
\BIBentrySTDinterwordspacing

\bibitem{pulpSdk}
\BIBentryALTinterwordspacing
``{PULP SDK}.''  \url{https://github.com/pulp-platform/pulp-sdk}
\BIBentrySTDinterwordspacing

\bibitem{cv32e40p}
\BIBentryALTinterwordspacing
``{OpenHW Group CORE-V CV32E40P RISC-V IP}.''
  \url{https://github.com/openhwgroup/cv32e40p}
\BIBentrySTDinterwordspacing

\bibitem{ibex}
\BIBentryALTinterwordspacing
``{Ibex RISC-V Core}.''  \url{https://github.com/lowRISC/ibex}
\BIBentrySTDinterwordspacing

\bibitem{zedboard}
\BIBentryALTinterwordspacing
``{ZedBoard}.''  \url{http://zedboard.org/product/zedboard}
\BIBentrySTDinterwordspacing

\bibitem{coremark}
\BIBentryALTinterwordspacing
``{CoreMark}.''  \url{https://www.eembc.org/coremark/}
\BIBentrySTDinterwordspacing

\bibitem{embenchIoT}
\BIBentryALTinterwordspacing
``{Embench: Open Benchmarks for Embedded Platforms}.''
  \url{https://github.com/embench/embench-iot}
\BIBentrySTDinterwordspacing

\bibitem{riscvBenchmarks}
\BIBentryALTinterwordspacing
``{Benchmarks for RISCV Processor}.''
  \url{https://github.com/ucb-bar/riscv-benchmarks}
\BIBentrySTDinterwordspacing

\bibitem{mibench}
\BIBentryALTinterwordspacing
``{MiBench}.''  \url{http://vhosts.eecs.umich.edu/mibench/}
\BIBentrySTDinterwordspacing

\end{thebibliography}


\begin{thebibliography}{10}
\providecommand{\url}[1]{#1}
\csname url@samestyle\endcsname
\providecommand{\newblock}{\relax}
\providecommand{\bibinfo}[2]{#2}
\providecommand{\BIBentrySTDinterwordspacing}{\spaceskip=0pt\relax}
\providecommand{\BIBentryALTinterwordstretchfactor}{4}
\providecommand{\BIBentryALTinterwordspacing}{\spaceskip=\fontdimen2\font plus
\BIBentryALTinterwordstretchfactor\fontdimen3\font minus
  \fontdimen4\font\relax}
\providecommand{\BIBforeignlanguage}[2]{{%
\expandafter\ifx\csname l@#1\endcsname\relax
\typeout{** WARNING: IEEEtranS.bst: No hyphenation pattern has been}%
\typeout{** loaded for the language `#1'. Using the pattern for}%
\typeout{** the default language instead.}%
\else
\language=\csname l@#1\endcsname
\fi
#2}}
\providecommand{\BIBdecl}{\relax}
\BIBdecl

\bibitem{abadi2009control}
M.~Abadi, M.~Budiu, {\'U}.~Erlingsson, and J.~Ligatti, ``{Control-Flow
  Integrity Principles, Implementations, and Applications},'' \emph{ACM
  Transactions on Information and System Security (TISSEC)}, vol.~13, no.~1,
  pp. 1--40, 2009.

\bibitem{abbasiChallengesDesigningExploitMitigations2019}
A.~Abbasi, J.~Wetzels, T.~Holz, and S.~Etalle, ``Challenges in designing
  exploit mitigations for deeply embedded systems,'' in \emph{2019 {{IEEE
  European Symposium}} on {{Security}} and {{Privacy}}
  ({{EuroS}}\&{{P}})}.\hskip 1em plus 0.5em minus 0.4em\relax {IEEE}, Jun.
  2019, pp. 31--46.

\bibitem{arora2006hardware}
D.~Arora, S.~Ravi, A.~Raghunathan, and N.~K. Jha, ``{Hardware-Assisted Run-Time
  Monitoring for Secure Program Execution on Embedded Processors},'' \emph{IEEE
  Transactions on Very Large Scale Integration (VLSI) Systems}, vol.~14,
  no.~12, pp. 1295--1308, 2006.

\bibitem{AsmDB}
G.~Ayers, N.~P. Nagendra, D.~I. August, H.~K. Cho, S.~Kanev, C.~Kozyrakis,
  T.~Krishnamurthy, H.~Litz, T.~Moseley, and P.~Ranganathan, ``{AsmDB}:
  Understanding and mitigating front-end stalls in warehouse-scale computers,''
  in \emph{Proceedings of the 46th International Symposium on Computer
  Architecture}.\hskip 1em plus 0.5em minus 0.4em\relax New York, NY, USA:
  Association for Computing Machinery, 2019, pp. 462--473.

\bibitem{burow2017control}
N.~Burow, S.~A. Carr, J.~Nash, P.~Larsen, M.~Franz, S.~Brunthaler, and
  M.~Payer, ``{Control-Flow Integrity: Precision, Security, and Performance},''
  \emph{ACM Computing Surveys (CSUR)}, vol.~50, no.~1, pp. 1--33, 2017.

\bibitem{carliniControlflowBendingEffectivenessControlflow2015}
N.~Carlini, A.~Barresi, M.~Payer, D.~Wagner, and T.~R. Gross, ``Control-flow
  bending: On the effectiveness of control-flow integrity,'' in
  \emph{24{\textsuperscript{th}} {{USENIX Security Symposium}} ({{USENIX
  Security}} 15)}.\hskip 1em plus 0.5em minus 0.4em\relax {USENIX Association},
  Aug. 2015, pp. 161--176.

\bibitem{checkowayReturnorientedProgrammingReturns2010}
S.~Checkoway, L.~Davi, A.~Dmitrienko, A.-R. Sadeghi, H.~Shacham, and
  M.~Winandy, ``Return-oriented programming without returns,'' in
  \emph{Proceedings of the 17th {{ACM Conference}} on {{Computer}} and
  Communications Security}.\hskip 1em plus 0.5em minus 0.4em\relax {New York,
  NY, USA}: {ACM}, Oct. 2010, pp. 559--572.

\bibitem{chiueh2001rad}
T.-c. Chiueh and F.-H. Hsu, ``{RAD: A compile-time solution to buffer overflow
  attacks},'' in \emph{Proceedings 21st International Conference on Distributed
  Computing Systems}.\hskip 1em plus 0.5em minus 0.4em\relax IEEE, 2001, pp.
  409--417.

\bibitem{christoulakis2016hcfi}
N.~Christoulakis, G.~Christou, E.~Athanasopoulos, and S.~Ioannidis, ``{HCFI}:
  Hardware-enforced control-flow integrity,'' in \emph{Proceedings of the Sixth
  ACM Conference on Data and Application Security and Privacy}, 2016, pp.
  38--49.

\bibitem{das2016fine}
S.~Das, W.~Zhang, and Y.~Liu, ``{A Fine-Grained Control Flow Integrity Approach
  Against Runtime Memory Attacks for Embedded Systems},'' \emph{IEEE
  Transactions on Very Large Scale Integration (VLSI) Systems}, vol.~24,
  no.~11, pp. 3193--3207, 2016.

\bibitem{davi2015hafix}
L.~Davi, M.~Hanreich, D.~Paul, A.-R. Sadeghi, P.~Koeberl, D.~Sullivan,
  O.~Arias, and Y.~Jin, ``{HAFIX}: Hardware-assisted flow integrity
  extension,'' in \emph{Proceedings of the 52{\textsuperscript{nd}} {{Annual
  Design Automation Conference}} - {{DAC}}'15}.\hskip 1em plus 0.5em minus
  0.4em\relax {New York, NY, USA}: {ACM}, 2015.

\bibitem{davi2014hardware}
L.~Davi, P.~Koeberl, and A.-R. Sadeghi, ``Hardware-assisted fine-grained
  control-flow integrity: Towards efficient protection of embedded systems
  against software exploitation,'' in \emph{Proceedings of the
  51{\textsuperscript{st}} {{Annual Design Automation Conference}} -
  {{DAC}}'14}.\hskip 1em plus 0.5em minus 0.4em\relax {San Francisco, CA, USA}:
  {IEEE}, Jun. 2014.

\bibitem{de2019fixer}
A.~De, A.~Basu, S.~Ghosh, and T.~Jaeger, ``{FIXER}: Flow integrity extensions
  for embedded {RISC-V},'' in \emph{2019 Design, Automation \& Test in Europe
  Conference \& Exhibition (DATE)}.\hskip 1em plus 0.5em minus 0.4em\relax
  IEEE, 2019, pp. 348--353.

\bibitem{de2017survey}
R.~de~Clercq and I.~Verbauwhede, ``{A survey of Hardware-based Control Flow
  Integrity (CFI)},'' \emph{arXiv preprint arXiv:1706.07257}, 2017.

\bibitem{ertl2003interpreters}
M.~A. Ertl and D.~Gregg, ``The structure and performance of efficient
  interpreters,'' \emph{Journal of Instruction-Level Parallelism}, vol.~5, pp.
  1--25, Nov. 2003.

\bibitem{gautschiNearthresholdRISCVCoreDSP2017}
M.~Gautschi, P.~D. Schiavone, A.~Traber, I.~Loi, A.~Pullini, D.~Rossi,
  E.~Flamand, F.~K. G{\"u}rkaynak, and L.~Benini, ``Near-threshold
  {{RISC}}-{{V}} core with {{DSP}} extensions for scalable {{IoT}} endpoint
  devices,'' \emph{IEEE Transactions on Very Large Scale Integration (VLSI)
  Systems}, vol.~25, no.~10, pp. 2700--2713, Oct. 2017.

\bibitem{goktasPositionindependentCodeReuseEffectiveness2018}
E.~G{\"o}ktas, B.~Kollenda, P.~Koppe, E.~Bosman, G.~Portokalidis, T.~Holz,
  H.~Bos, and C.~Giuffrida, ``Position-independent code reuse: On the
  effectiveness of {{ASLR}} in the absence of information disclosure,'' in
  \emph{2018 {{IEEE European Symposium}} on {{Security}} and {{Privacy}}
  ({{EuroS}}\&{{P}})}.\hskip 1em plus 0.5em minus 0.4em\relax {IEEE}, Apr.
  2018, pp. 227--242.

\bibitem{gonzalvezCaseIndirectJumpsSecure2019}
A.~Gonzalvez and R.~Lashermes, ``A case against indirect jumps for secure
  programs,'' in \emph{Proceedings of the 9th {{Workshop}} on {{Software
  Security}}, {{Protection}}, and {{Reverse Engineering}}}.\hskip 1em plus
  0.5em minus 0.4em\relax {New York, NY, USA}: {ACM}, Dec. 2019.

\bibitem{he2017no}
W.~He, S.~Das, W.~Zhang, and Y.~Liu, ``{No-Jump-into-Basic-Block: Enforce Basic
  Block CFI on the Fly for Real-world Binaries},'' in \emph{Proceedings of the
  54{\textsuperscript{th}} {{Annual Design Automation Conference}} -
  {{DAC}}'17}.\hskip 1em plus 0.5em minus 0.4em\relax ACM, 2017.

\bibitem{IntelCET}
{Intel}, ``{Control-flow Enforcement Technology Specification},'' May 2019.

\bibitem{kayaalp2012efficiently}
M.~Kayaalp, M.~Ozsoy, N.~A. Ghazaleh, and D.~Ponomarev, ``{Efficiently Securing
  Systems from Code Reuse Attacks},'' \emph{IEEE Transactions on Computers},
  vol.~63, no.~5, pp. 1144--1156, 2012.

\bibitem{kayaalp2013scrap}
M.~Kayaalp, T.~Schmitt, J.~Nomani, D.~Ponomarev, and N.~Abu-Ghazaleh, ``{SCRAP:
  Architecture for Signature-Based Protection from Code Reuse Attacks},'' in
  \emph{2013 IEEE 19th International Symposium on High Performance Computer
  Architecture (HPCA)}.\hskip 1em plus 0.5em minus 0.4em\relax IEEE, 2013, pp.
  258--269.

\bibitem{levySmashingStackFunProfit1996}
\BIBentryALTinterwordspacing
E.~Levy, ``Smashing the stack for fun and profit,'' \emph{Phrack}, no.~49, Nov.
  1996. [Online]. Available: \url{http://phrack.org/issues/49/14.html}
\BIBentrySTDinterwordspacing

\bibitem{rahmatian2012hardware}
M.~Rahmatian, H.~Kooti, I.~G. Harris, and E.~Bozorgzadeh, ``{Hardware-Assisted
  Detection of Malicious Software in Embedded Systems},'' \emph{IEEE Embedded
  Systems Letters}, vol.~4, no.~4, pp. 94--97, 2012.

\bibitem{randalIdealRealRevisitingHistory2020}
A.~Randal, ``The ideal versus the real: Revisiting the history of virtual
  machines and containers,'' \emph{ACM Computing Surveys}, vol.~53, no.~1, pp.
  5:1--5:31, Feb. 2020.

\bibitem{samuelgrossProjectZeroJITSploitationIII2020}
\BIBentryALTinterwordspacing
{Samuel Gro\ss}, ``Project {{Zero}}: {{JITSploitation III}}: Subverting control
  flow,'' Sep. 2020. [Online]. Available:
  \url{https://googleprojectzero.blogspot.com/2020/09/jitsploitation-three.html}
\BIBentrySTDinterwordspacing

\bibitem{schiavoneSlowSteadyWinsRace2017}
P.~D. Schiavone, F.~Conti, D.~Rossi, M.~Gautschi, A.~Pullini, E.~Flamand, and
  L.~Benini, ``Slow and steady wins the race? {{A}} comparison of
  ultra-low-power {{RISC}}-{{V}} cores for {{Internet}}-of-{{Things}}
  applications,'' in \emph{2017 27th {{International Symposium}} on {{Power}}
  and {{Timing Modeling}}, {{Optimization}} and {{Simulation}}
  ({{PATMOS}})}.\hskip 1em plus 0.5em minus 0.4em\relax {Thessaloniki, Greece}:
  {IEEE}, Sep. 2017.

\bibitem{scutExploitingFormatStringVulnerabilities2001}
{scut}, ``Exploiting format string vulnerabilities,'' {Team TESO}, Tech. Rep.,
  Sep. 2001.

\bibitem{shachamGeometryInnocentFleshBone2007}
H.~Shacham, ``The geometry of innocent flesh on the bone: Return-into-libc
  without function calls (on the x86),'' in \emph{Proceedings of the 14th {{ACM
  Conference}} on {{Computer}} and Communications Security}.\hskip 1em plus
  0.5em minus 0.4em\relax {New York, NY, USA}: {ACM}, Oct. 2007, pp. 552--561.

\bibitem{songSoKSanitizingSecurity2019}
D.~Song, J.~Lettner, P.~Rajasekaran, Y.~Na, S.~Volckaert, P.~Larsen, and
  M.~Franz, ``{{SoK}}: Sanitizing for security,'' in \emph{2019 {{IEEE
  Symposium}} on {{Security}} and {{Privacy}} ({{S\&P}})}.\hskip 1em plus 0.5em
  minus 0.4em\relax {IEEE}, May 2019, pp. 1275--1295.

\bibitem{sullivan2016strategy}
D.~Sullivan, O.~Arias, L.~Davi, P.~Larsen, A.-R. Sadeghi, and Y.~Jin,
  ``Strategy without tactics: Policy-agnostic hardware-enhanced control-flow
  integrity,'' in \emph{Proceedings of the 53{\textsuperscript{rd}} {{Annual
  Design Automation Conference}} - {{DAC}}'16}.\hskip 1em plus 0.5em minus
  0.4em\relax {ACM}, 2016.

\bibitem{excec_repo}
M.~Telesklav and S.~Tauner, ``{EXCEC} repositories,'' 2021, {URL to be
  announced}.

\bibitem{theodorides2017breaking}
M.~Theodorides and D.~Wagner, ``Breaking active-set backward-edge {CFI},'' in
  \emph{2017 IEEE International Symposium on Hardware Oriented Security and
  Trust (HOST)}.\hskip 1em plus 0.5em minus 0.4em\relax IEEE, 2017, pp. 85--89.

\bibitem{wenzl2019hack}
M.~Wenzl, G.~Merzdovnik, J.~Ullrich, and E.~Weippl, ``{From Hack to Elaborate
  Technique—A Survey on Binary Rewriting},'' \emph{ACM Computing Surveys
  (CSUR)}, vol.~52, no.~3, pp. 1--37, 2019.

\bibitem{wilkenContinuousSignatureMonitoringLowcost1990}
K.~Wilken and J.~P. Shen, ``Continuous signature monitoring: Low-cost
  concurrent detection of processor control errors,'' \emph{IEEE Transactions
  on Computer-Aided Design of Integrated Circuits and Systems}, vol.~9, no.~6,
  pp. 629--641, Jun. 1990.

\end{thebibliography}

\bibliographystyleLinks{IEEEtranUrldate}
\bibliographyLinks{}

\end{document}